\documentclass{article}

\usepackage{microtype}
\usepackage{graphicx}
\usepackage{subfigure}
\usepackage{booktabs} 

\usepackage{amsmath,amsfonts,bm}

\def\eqref#1{equation~\ref{#1}}

\def\1{\bm{1}}

\DeclareMathAlphabet{\mathsfit}{\encodingdefault}{\sfdefault}{m}{sl}
\SetMathAlphabet{\mathsfit}{bold}{\encodingdefault}{\sfdefault}{bx}{n}

\def\gC{{\mathcal{C}}}

\def\gG{{\mathcal{G}}}

\def\gL{{\mathcal{L}}}

\def\gX{{\mathcal{X}}}

\def\sX{{\mathbb{X}}}

\def\sZ{{\mathbb{Z}}}

\newcommand{\E}{\mathbb{E}}
\newcommand{\dd}{\mathrm{d}}

\newcommand{\R}{\mathbb{R}}

\def\dimx{{d_\sX}}
\def\dimz{{d_\sZ}}
\usepackage{hyperref}
\usepackage{pifont}

\usepackage[accepted]{icml2024}

\usepackage{amsmath}
\usepackage{amssymb}
\usepackage{mathtools}
\usepackage{amsthm}

\usepackage {makecell} 
\usepackage{multirow}

\usepackage[capitalize,noabbrev]{cleveref}

\theoremstyle{plain}
\newtheorem{theorem}{Theorem}[section]

\newtheorem{lemma}[theorem]{Lemma}
\newtheorem{corollary}[theorem]{Corollary}
\theoremstyle{definition}
\newtheorem{definition}[theorem]{Definition}

\theoremstyle{remark}

\usepackage{algpseudocode}
\usepackage[textsize=tiny]{todonotes}

\icmltitlerunning{UniCorn: A Unified Contrastive Learning Approach for Multi-view Molecular Representation Learning}
\begin{document}



\twocolumn[
\icmltitle{UniCorn: A Unified Contrastive Learning Approach for Multi-view Molecular Representation Learning}



\icmlsetsymbol{equal}{*}

\begin{icmlauthorlist}
\icmlauthor{Shikun Feng}{equal,scht}
\icmlauthor{Yuyan Ni}{equal,scha,schb}
\icmlauthor{Minghao Li}{schc}
\icmlauthor{Yanwen Huang}{schd}
\icmlauthor{Zhi-Ming Ma}{scha}
\icmlauthor{Wei-Ying Ma}{scht}
\icmlauthor{Yanyan Lan}{scht}
\end{icmlauthorlist}

\icmlaffiliation{scht}{Institute for AI Industry Research (AIR), Tsinghua University}
\icmlaffiliation{scha}{Academy of Mathematics and Systems Science, Chinese Academy of Sciences }
\icmlaffiliation{schb}{University of Chinese Academy of Sciences}
\icmlaffiliation{schc}{Beijing Institute of Genomics, Chinese Academy of Sciences}
\icmlaffiliation{schd}{Department of Pharmaceutical Science, Peking University}

\icmlcorrespondingauthor{Yanyan Lan}{lanyanyan@air.tsinghua.edu.cn}

\vskip 0.3in
]



\printAffiliationsAndNotice{\icmlEqualContribution} 


\begin{abstract}
Recently, a noticeable trend has emerged in developing pre-trained foundation models in the domains of CV and NLP. However, for molecular pre-training, there lacks a universal model capable of effectively applying to various categories of molecular tasks, since existing prevalent pre-training methods exhibit effectiveness for specific types of downstream tasks. Furthermore, the lack of profound understanding of existing pre-training methods, including 2D graph masking, 2D-3D contrastive learning, and 3D denoising, hampers the advancement of molecular foundation models. In this work, we provide a unified comprehension of existing pre-training methods through the lens of contrastive learning. Thus their distinctions lie in clustering different views of molecules, which is shown beneficial to specific downstream tasks. To achieve a complete and general-purpose molecular representation, we propose a novel pre-training framework, named UniCorn, that inherits the merits of the three methods, depicting molecular views in three different levels. SOTA performance across quantum, physicochemical, and biological tasks, along with comprehensive ablation study, validate the universality and effectiveness of UniCorn. 
\end{abstract}

\begin{figure*}[t]
\begin{center}
\centerline{\includegraphics[width=1\textwidth]{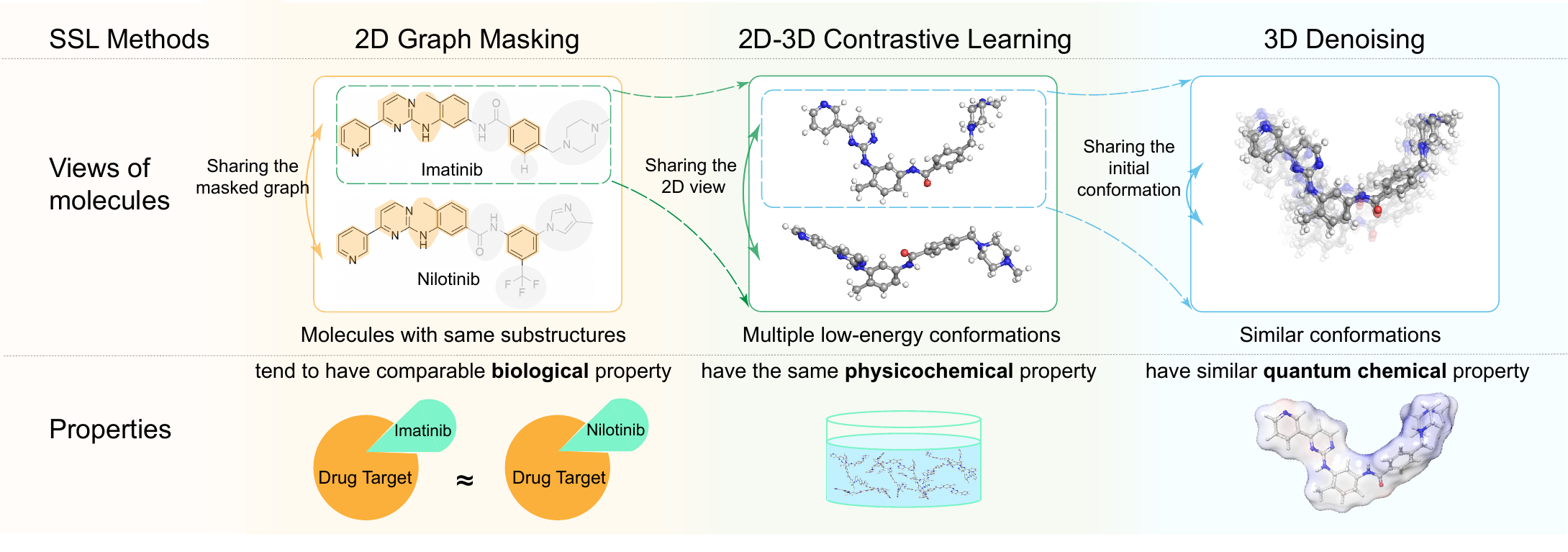}}
\vskip -10pt
\caption{ Correspondence between self-supervised learning (SSL) methods, views of molecules, and molecular properties. 
Different SSL methods cluster the molecular representations based on different levels of similarity (section~\ref{sec:theory unify methods}). These clustering patterns align with the characteristics of properties at different scales (section~\ref{sec:relation between method&task}).
}
\label{fig:motivation}
\end{center}
\vskip -28pt
\end{figure*}
\vspace{-15pt}

\section{Introduction}
\label{submission}
Molecular representation learning is pivotal across diverse drug discovery tasks.
One important application is molecular property prediction which is promising to enable high-throughput screening of molecules with desirable properties.
Following the pre-training methods in natural language processing (NLP) and computer vision (CV), a range of molecular pre-training methods has emerged to address the challenge of limited labeled molecular data. Specifically, they perform self-supervised learning (SSL) on a large number of unlabeled data, and then fine-tune on specific kinds of property prediction tasks with labeled data. Existing SSL methods can be mainly classified into three categories. The first category involves 2D graph masking~\citep{hu2020strategies, hou2022graphmae, rong2020self, xia2022mole}, where random parts of the molecular graph are masked and the model is pre-trained to reconstruct them. The second category comprises 2D-3D contrastive learning~\citep{stark20223d, liu2021pre, liu2023group, li2022geomgcl} that aligns the representations of 3D conformations and that of their corresponding 2D graph. The third category is 3D denoising~\citep{zaidi2022pre, feng2023fractional, luo2022one, liu2022molecular,ni2024sliced} that adds noise to the conformation and trains the model to predict the noise.


Recently, foundation models, such as ChatGPT~\citep{chatgpt}, GPT4~\citep{gpt4}, SAM~\citep{kirillov2023segany}, Unified-IO~\citep{lu2022unified} 
, have caused a revolutionary shift to the field of Artificial Intelligence, which are pre-trained on large-scale data and are adopted to a broad range of downstream tasks~\citep{li2023multimodal}. However, there still lacks a unified model in the molecular domain, that can learn a universal representation and effectively be applied to various property prediction tasks. The challenge firstly lies in that the relationship between the existing SSL methods is still under study.
Moreover, existing methods are unbalanced for various downstream tasks. Generally, 3D denoising methods favor quantum chemical property prediction, while 2D graph masking and 2D-3D contrastive learning prefer biological and physicochemical property prediction. This phenomenon, also manifested in section~\ref{sec:main exp}, is hardly discussed by previous studies and the causes are still unclear.
Furthermore, the relationship between the existing SSL methods is still under study, posing a challenge in leveraging their strengths to create a universal model effective for all three types of property prediction tasks.

 To tackle these challenges, we provide a unified understanding of the three SSL methods through contrastive learning in section~\ref{sec:theory unify methods}. In particular, 2D graph masking, 2D-3D contrastive
learning, and 3D denoising can be comprehended as contrastive learning with masking, sampling multiple conformations of the same molecule, and noise addition as augmented views, respectively. In the derivation, we first of all summarize masking and denoising as reconstructive methods. Then we find the contrastive loss and the reconstructive loss are mutually upper and lower bounded by each other, indicating minimizing one loss can guarantee the other loss to be small under certain conditions and regularization. Lastly, we interpret the representation of contrastive learning as clustering different views.
Therefore, the three SSL methods lead to \textbf{clustering patterns in molecular representation space at different granularity}. Importantly, clustering patterns at different granularity can exist concurrently, suggesting that the three pre-training methods are compatible, resulting in a multi-grained representation. 
 Moreover,  we reveal that the multi-grained clustering patterns correspond to the inductive biases of different kinds of molecular properties, as shown in Figure~\ref{fig:motivation}. 
Firstly, 3D denoising aligns the representations of closely resembled conformations, in accordance with the characteristics of quantum chemical properties. Secondly, 2D-3D contrastive learning learns the invariance of multiple low-energy conformations of one molecule, corresponding to the characteristics of physicochemical properties. Thirdly, 2D graph masking clusters molecules with the same fragments, capturing characteristics of biological properties. Elaborate elucidation is provided in section~\ref{sec:relation between method&task}.  \textbf{This correspondence highlights the necessity to combine the three pre-training methods in order to achieve a universal representation.}

Consequently, we propose UniCorn, a \textbf{uni}fied molecular pre-training framework via \textbf{co}ntrastive lea\textbf{rn}ing, to learn multi-view molecular representations by amalgamating the strengths of existing methods into a unified pre-training framework, naturally capable of tackling quantum chemical, physicochemical, and biological properties.
UniCorn takes both 2D molecular graphs and 3D molecular conformations as input, with tailored self-supervised strategies for each data type.
For 2D graphs, we utilize fragments, recognized as a kind of chemical semantic component, as masking units to mask the 2D graphs and subsequently recover them. 
Regarding the 3D conformations, we employ torsion augmented denoising, which first augments rotatable torsions of the molecule to sample multiple chemically plausible 3D conformations, then perturbs the coordinate of the augmented molecule and predicts associated coordinate noise. 
Finally, cross-modal contrastive learning is introduced to align the representations of multiple 3D conformations generated by torsion augmented denoising to their shared 2D representation and further distill knowledge from 2D to 3D. 

Experimental results show that UniCorn is not confined to achieving merely comparable results on tasks where existing methods excel but also consistently surpasses them across all three types of tasks. This not only demonstrates the compatibility of each pre-training module in UniCorn but also highlights their complementarity in achieving a universal molecular representation.
Ablation study and visualization further validate our comprehension of SSL methods and their correlation to downstream tasks.
Our contributions are summarized as follows:
\vspace{-2pt}
\begin{itemize}
\item  To our best knowledge, we are the first to systematically summarize existing molecular pre-training methods and conduct a thorough analysis of their associations with various downstream molecular tasks.
\vspace{-2pt}
\item Theoretically, we reveal the connection between reconstructive and contrastive methods and comprehend them in a unified perspective through contrastive learning and representation clustering.
\vspace{-2pt}
\item  Practically, we present UniCorn, a unified pre-training framework designed to learn hierarchical molecular representations applicable to a wide array of downstream tasks.
\vspace{-2pt}
\item Through exhaustive experiments on 
physicochemical and biological tasks in MoleculeNet and quantum tasks including QM9, MD17, and MD22, UniCorn surpasses existing molecular pre-training methods, showcasing its superior performance.
\end{itemize}


\section{Unifying Reconstructive and Contrastive Methods}\label{sec:theory unify methods}

\subsection{Abstraction and Formulation}
In general, denoising and masking can be concluded as the reconstructive method that aims to reconstruct the original input from the perturbed version containing noise or masked elements. The introduction of noise or masking provides an alternative view of the original input, effectively serving as a form of data augmentation.

Contrastive learning, on the other hand, aims to align representations of different views of the input. To this end, several different families of methods have been raised, as summarized in \citet{2023cookbook}, including the deep metric learning family that usually involves negative samples and a symmetric structure~\cite{chen2020simclr,oord2018infonce}, the self-distillation family that adopts a predictor to map between representations~\cite{grill2020byol,chen2021simsiam,caron2021dino}, and the canonical correlation analysis family that additionally regularize the covariance of the representations~\cite{bardes2021vicreg}. To show the clear correspondence between contrastive and reconstructive methods, we discuss the SimSiam like contrastive method~\cite{chen2021simsiam} in the self-distillation family. The unification and equivalence between the three contrastive families are discussed in \citet{tao2022equi1,garrido2022duality}.

\begin{figure}[t]
\begin{center}
\centerline{\includegraphics[width=0.9\columnwidth]{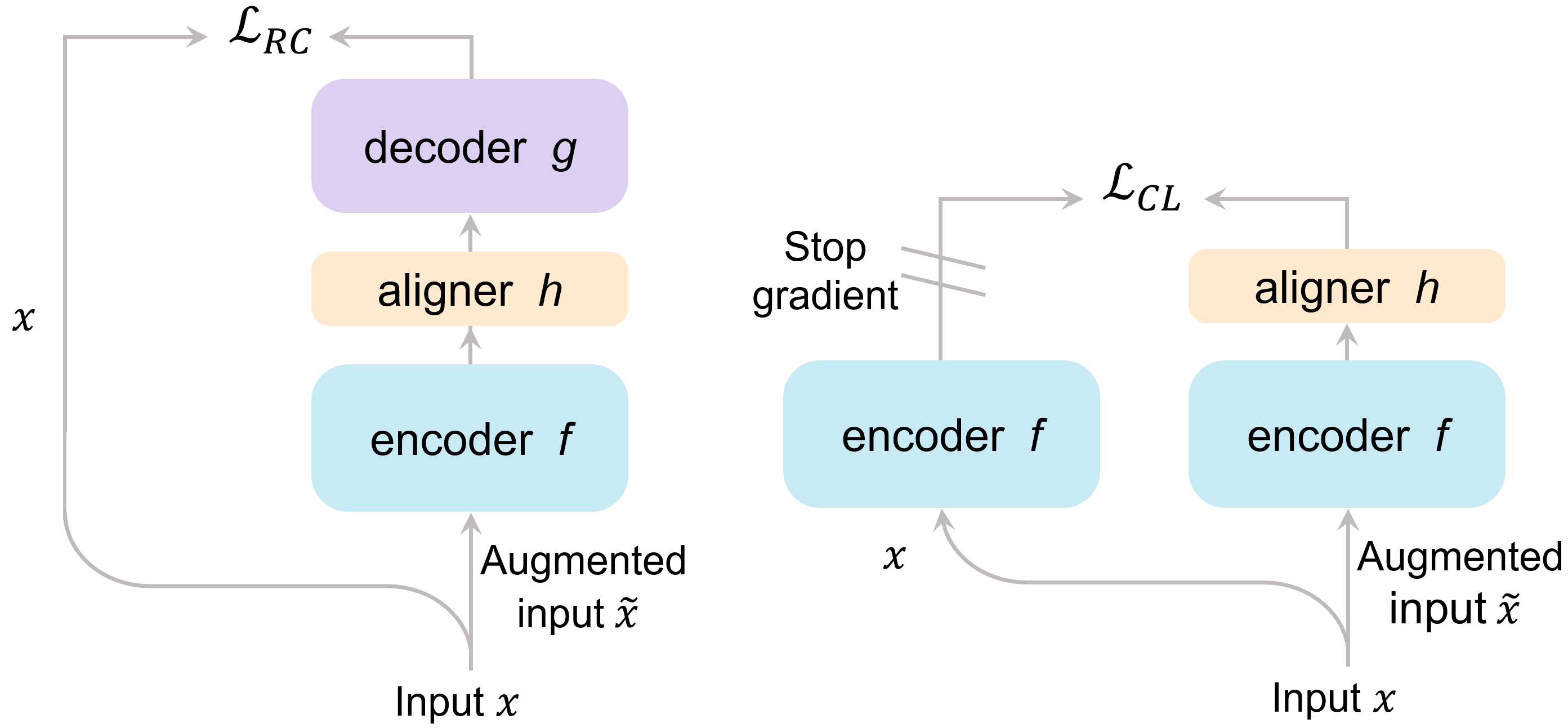}}
\caption{A unified perspective of reconstructive and contrastive methods.}
\label{fig:rccl}
\end{center}
\vskip -30pt 
\end{figure}

To formalize the learning objectives, we introduce $\sX\subseteq \R^\dimx$, $\tilde{\sX}\subseteq \R^\dimx$ to denote the sets comprising raw input data and the augmented data, respectively. We use $p(x)$ and $p(\tilde{x}|x)$ to denote the distribution of input data $x\in\sX$ and the probability of generating an augmentation $\tilde{x}\in \tilde{\sX}$ from $x$. The representations are situated within a representation space in $\R^\dimz$. 
To facilitate a unified perspective, we define an encoder function $f_\theta:\sX\rightarrow \R^\dimz$, an aligner function $h_\psi:\R^\dimz\rightarrow \R^\dimz$ and a continuously differentiable decoder function $g_\phi:\R^\dimz\rightarrow \sX$, for extracting primary features from the input, aligning representations, and reconstructing the input from the representations, respectively. Here, $\theta$, $\psi$ and $\phi$ denote learnable parameters associated with these functions. As a result, the reconstructive loss and contrastive loss can be expressed as:
 \begin{equation}        \gL_{\text{RC}}=\mathbb{E}_{p(x)}\mathbb{E}_{p(\tilde{x}|x)}||g_\phi(h_\psi(f_\theta(\tilde{x})))-x||_2,
    \end{equation}
    \begin{equation}\label{eq:cl loss}
        \gL_{\text{CL}}=\mathbb{E}_{p(x)}\mathbb{E}_{p(\tilde{x}|x)}||h_\psi(f_\theta(\tilde{x}))-\mathcal{SG}(f_\theta(x))||_2,
    \end{equation}
    where $||\cdot||_2$ represents the $L^2$ norm of vectors. While the squared $L^2$ norm is conventionally employed, we opt for the $L^2$ norm for the sake of brevity and clarity in the proof. This choice does not impact the discussion on loss functions, as the gradient vectors of these functions align in the same direction. The notation $\mathcal{SG}$ signifies stop gradient. An illustration of contrastive and reconstructive methods in a unified perspective is shown in Figure~\ref{fig:rccl}.

\subsection{Mutual Upper and Lower Bounds}
In this section, we demonstrate that, under certain conditions and regularization techniques, reconstructive and contrastive losses can serve as mutual upper and lower bounds for each other. This implies that minimizing one loss inherently results in the optimization of the other.
\begin{theorem}[Relations between reconstructive and contrastive loss]\label{thm: mutual bound}
We introduce an additional loss aimed at regularizing the decoder to approximate the inverse of the encoder.
\begin{equation}\label{eq:thm CL>RC}         \gL_{\text{reg}}=\mathbb{E}_{p(x)}||g_\phi( \mathcal{SG}(f_\theta(x)) )-x||_2
    \end{equation}
When $\lambda_{\text{max}}$ and $\lambda_{\text{min}}$ are non-zero, we derive the following conclusions:\\
    \begin{minipage}[b]{0.05\textwidth}
       \ding{172}
    \end{minipage}
    \begin{minipage}[b]{0.425\textwidth}
       \begin{equation}\label{eq:thm RC>CL}
             \gL_{\text{CL}}+ \frac{1}{\lambda_{\text{max}}}\gL_{\text{reg}}
             \geq \frac{1}{\lambda_{\text{max}}} \gL_{\text{RC}} 
        \end{equation}
    \end{minipage}   
        indicating that updating the encoder and aligner via contrastive learning and updating the decoder by the regularization loss, guarantees a small reconstructive loss. \\
    \begin{minipage}[b]{0.05\textwidth}
       \ding{173}
    \end{minipage}
    \begin{minipage}[b]{0.425\textwidth}
       \begin{equation}
         \gL_{\text{RC}}+ \gL_{\text{reg}}\geq \lambda_{\text{min}} \gL_{\text{CL}},
    \end{equation} 
    \end{minipage}                
    indicating that updating the entire network through reconstructive learning and selectively partitioning the decoder by regularization loss guarantees a small contrastive loss. Kindly note that the partition of the encoder and decoder from the entire network does not affect $\gL_{\text{RC}}$, however it is crucial for contrastive learning. Therefore, the regularization term can be optimized by finding the optimal partition of the encoder and decoder from the entire network. And the contrastive loss is defined by the partitioned encoder and aligner.
\end{theorem}
Here, $\lambda_{\text{max}}$ and $\lambda_{\text{min}}$ are constants defined in the proof. 
The proof and discussions regarding the condition are elaborated in appendix~\ref{app: sec proof&analyse}.

\subsection{Clustered Representations}
Previous studies~\cite{yimingyang2021,2022hiddenuniform} have revealed the connection between clustering and symmetric contrastive learning methods, such as SimCLR~\cite{chen2020simclr} and VICReg~\cite{bardes2021vicreg}.
However, the contrastive loss defined in \eqref{eq:cl loss} exhibits asymmetry: the two representations to be aligned comprise one as the direct output of the encoder while the other passes through an aligner.
In essence, \eqref{eq:cl loss} resembles the SimSiam contrastive loss up to the square: $\gL_{\text{CL}}^{(\text{SimSiam})}=\E_{p(x)}\E_{p(\tilde{x}|x)}||h_\psi(f_\theta(\tilde{x}))-\mathcal{SG}(f_\theta(x))||_2^2$,  where the representations are normalized $||f_\theta(\cdot)||_2^2=||h_\psi(f_\theta(\cdot))||_2^2=1$. An analogy to k-means clustering of the SimSiam contrastive loss is discussed in ~\citet{chen2021simsiam}, where they hypothesized $ h_\psi (f_\theta(\tilde{x}))\approx\E_{p(\tilde{x}|x)} f_\theta(\tilde{x}) \approx f_\theta(x) $. We provide a more explicit relationship between clustering and the SimSiam loss without incorporating an additional hypothesis.
\begin{theorem}\label{thm:culster}    Minimizing the contrastive loss guarantees clustering of the augmentation representations corresponding to the same raw input data.
    \begin{equation}
    \begin{aligned}
        \gL_{\text{CL}}^{(\text{SimSiam})}\geq \gL_{\text{cluster}}. 
    \end{aligned}
    \end{equation}
     $\gL_{\text{cluster}}  \triangleq \E_{p(x)}\E_{p(\tilde{x}|x)} ||h_\psi(f_\theta(\tilde{x}))- \E_{p(\tilde{x}|x)} h_\psi(f_\theta(\tilde{x}))||_2^2$ describes the mean distance between the samples in the cluster and the cluster center.
\end{theorem}
The theorem indicates minimizing the SimSiam contrastive loss guarantees a small clustering loss. The proof is provided in section~\ref{app sec: proof cluster}. Consequently, we can comprehend the reconstructive and contrastive methods from a unified perspective, as they both cluster different views of the input. 
\section{Delving into the Task-specific Preferences of Existing Methods}\label{sec:relation between method&task}
 The previously introduced theorems have provided a unified comprehension of all three molecular SSL methods via contrastive learning and clustering of the representations. In molecular pre-training, the pivotal distinction among these methods lies in data augmentation approaches, which emphasize different views of molecules. Overall, we find that the clustered representations of each SSL method tend to align with a specific category of downstream tasks, which is commonly regarded as desirable for better generalization \cite{tian2020makes,wang2022ladder,haochen2021spectralcl,yimingyang2021}, as well as efficient few-shot transfer learning \cite{galanti2021fewshot,galanti2022transfer}.

\subsection{2D Graph Masking}\label{sec:3.1}
In our comprehension of 2D graph masking, the data augmentation is randomly masking atoms, edges, or fragments in the molecular graph~\cite{hu2020strategies, hou2022graphmae,rong2020self,feng2023unimap}. When similar molecules share a majority of the same substructures, their augmentation can overlap by masking non-identical regions and retaining shared substructures. As a result, these similar molecules are connected in the augmentation graph and can be clustered during pre-training~\cite{wang2022ladder}. 

This clustering pattern is consistent with a fundamental principle in medicinal chemistry: molecules with similar substructures often exhibit comparable biological activities~\cite{johnson1990similarity1,Dean1995similarity2,willett1998similarity3}. An example in point is the FDA-approved breakthrough drugs, imatinib \cite{druker2001imatinib} and nilotinib \cite{breccia2010nilotinib}, shown in Figure \ref{fig:motivation}. Sharing identical fragments, these two drugs are both orally available, potent, small-molecule inhibitors targeting breakpoint cluster region-Abelson (BCR-ABL) with some comparable clinical outcomes, including the overall frequency of adverse events, ten-year progression-free survival rates, and ten-year overall survival~\cite{kantarjian2021long}. Their shared substructures contribute greatly to their similar biological properties both in vitro and in vivo~\cite{manley2005advances}.


\subsection{2D-3D Contrastive Learning}
Recent methods advocate the alignment of 2D and 3D representations of the same molecule during pre-training to effectively leverage both 2D and 3D information~\citep{stark20223d, liu2021pre, liu2023group, li2022geomgcl}, Among them using multiple conformations of the same molecule tends to achieve higher performance~\citep{stark20223d, liu2021pre}. According to a corollary of 
Theorem~\ref{thm:culster} in appendix~\ref{app sec corollary}, the cross-modal contrastive learning clusters diverse conformations of the same molecule in the 3D representation space.

This clustering pattern also aligns with the current understanding in the field of chemistry that certain physicochemical properties of the system are resilient to conformational changes of a single molecule within. 
There is an overall statistical distribution describing the system with molecules interconverting between different conformations \cite{allen1996conformer,brameld2008conformation}. As illustrated in Figure~\ref{fig:motivation}, many physicochemical properties reflect the features of such distribution, including water solubility and octanol/water distribution coefficient, i.e., ESOL and Lipophilicity tasks in MoleculeNet dataset~\citep{wu2018moleculenet} respectively, capturing the averaged behavior of molecular conformations rather than individual molecular conformation \cite{le2023conformations}.
%


\begin{figure*}[th]
\begin{center}
\centerline{\includegraphics[width=1.0\textwidth]{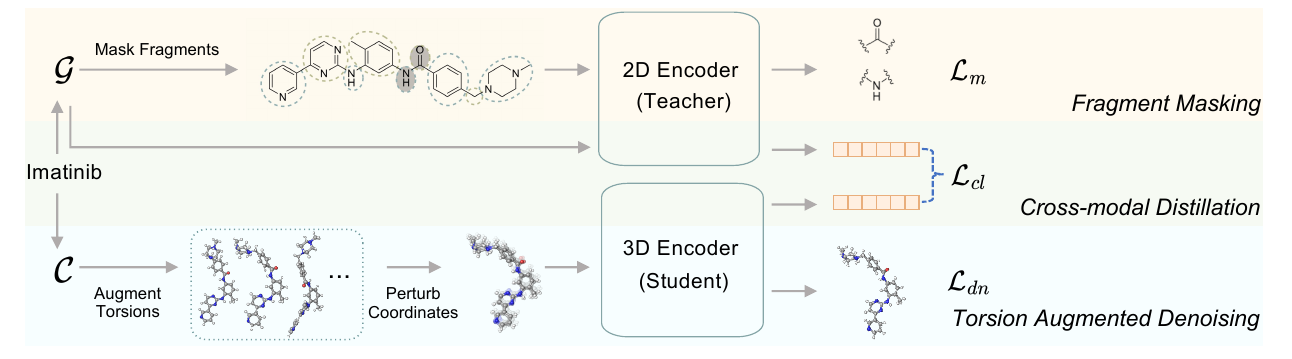}}
\caption{The UniCorn architecture we have ultimately reached, after exploring the association between each pre-training method and downstream tasks, with the goal of approaching unified molecular representations. The \textbf{Top} illustrates the Fragment Masking Module, wherein a 2D molecular graph is masked by fragments and subsequently recovered. The \textbf{Bottom} showcases the Torsion Augmented Denoising Module. This module operates in two steps: initially augmenting 3D conformers by perturbing rotatable torsions, and then introducing Gaussian coordinate noise for denoising. Finally, the \textbf{Middle} introduces the Cross-modal Distillation Module, responsible for distilling knowledge from 2D to 3D to achieve a hierarchical molecular representation.}
\label{fig:model}
\end{center}
\vskip -20pt
\end{figure*}

\subsection{3D Denoising} 
In our comprehension of 3D denoising, the data augmentation is adding Gaussian coordinate noise to the molecular conformation. As a result, 3D denoising clusters the representations of the noisy conformations around their equilibrium input. As demonstrated in Figure~\ref{fig:motivation}, this clustering pattern benefits quantum chemical properties as the small scale of perturbation ensures that the augmented conformations share similar properties to the raw conformation. This is because the quantum chemical properties are generally influenced by the spatial distribution of charges in molecules~\cite{kauzmann2013quantumbook}
, and thus undergo slight alterations when conformational changes are minor. For instance, the dipole moment in QM9 \citep{ramakrishnan2014quantum, ruddigkeit2012enumeration} dataset, a common quantum chemistry property indicating the separation of positive and negative charges in a molecule, is determined by the sum of products of charge magnitudes and their location vectors~\cite{LANDAU197589}. 
For similar conformations, meaning their atomic nuclei positions are largely unchanged, the spatial arrangement of the electron cloud exhibits negligible changes, resulting in similar dipole moments.

\section{UniCorn}
To learn a complete representation of molecules aimed at capturing hierarchical clustering patterns essential for covering molecular task preferences, we propose UniCorn which contains fragment masking, torsion augmented denoising, and cross-modal distillation modules that capture different scales of molecular information, as illustrated in Figure~\ref{fig:model}.
\subsection{Preliminary Notations}
A molecule can be either represented as a 2D molecular graph $\gG = (\mathcal{V}, \mathcal{E})$ or 3D conformation $\gC = (\mathcal{V}, \gX)$, where vertexes $\mathcal{V}$ signify the atom types, edges $\mathcal{E}$ denote the chemical bonds, and $\mathcal{X}$ signify the Cartesian coordinates.
\subsection{Fragment Masking Module}
To cluster molecules with the same chemically plausible substructures, we conduct Masked Fragment Modeling (MFM) on the molecular 2D graph. 
A fragment is typically a small segment of a large molecule capable of existing independently while preserving meaningful chemical properties or biological functionalities~\citep{erlanson2016twenty,murray2009rise,erlanson2004fragment}. 
Unlike previous methods that mask atoms, edges, or tokens, We treat fragments as the basic unit for masking better fitting the characteristic of biological tasks discussed in section~\ref{sec:3.1}.
Specifically, we utilize the BRICS~\citep{degen2008art} algorithm to break down the molecule into a set of fragments denoted as $\mathbf{s}$. We randomly mask a certain number of fragments, represented as $\mathbf{s}_{m}$, based on the mask ratio $m$. We then predict each atom type within the masked fragments. The masked fragment loss is defined by the following equation:

\begin{equation}
    \mathcal{L}_m = -\mathbb{E}_{\mathcal{G},\mathcal{G}_{\backslash m}}\sum_{\mathbf{s}_{m} } \log \left[p_{(g_m, f^{2d})}\left(\mathbf{s}_{m} |\mathcal{G}_{\backslash m} \right)\right]
\end{equation}
where $f^{2d}$ denotes the 2D graph encoder, $g_m$ represents the MLP head for MFM prediction, $p_{(g_m, f^{2d})}$ stands for the predicted probability of masked fragments, and $\mathcal{G}_{\backslash m}$ signifies the remaining part of the graph after masking.

\subsection{Torsion Augmented Denoising Module}
To cluster closely resembling conformations, we utilize the denoising pre-training strategy on 3D conformations. 
Unlike traditional denoising~\cite{zaidi2022pre} whose unperturbed conformations are equilibriums $\mathcal{C} = (\mathcal{V}, \mathcal{X})$ that come from the dataset, we innovatively introduce a torsion augmentation step before denoising to provide diverse conformations of the same molecular graph for the cross-modal distillation module which is elaborated in the next section.
Specifically, we perturb the torsion angles of the rotatable single bonds in equilibrium conformation to obtain multiple conformations for the molecule, denoted as $\mathcal{C}_a = (\mathcal{V}, \mathcal{X}_a)$. Subsequently, we adopt traditional denoising~\cite{zaidi2022pre} on $\mathcal{C}_a$. We apply Gaussian noise to the coordinates of each atom 
 and yield conformation $\tilde{\mathcal{C}} = (\mathcal{V}, \tilde{\mathcal{X}})$. The denoising loss is defined as follows:
\begin{equation}
\mathcal{L}_{dn} = \mathbb{E}_{\tilde{\mathcal{C}},\mathcal{C}_a}||g_{d}(f^{3d}(\tilde{\mathcal{C}})) - (\tilde{\mathcal{X}} - \mathcal{X}_a)||_{2}^{2}.
\end{equation}
Here, $f^{3d}$ is the 3D encoder, and $g_{d}$ represents the MLP head utilized for predicting the coordinate noise.
Please note that the target of denoising is to predict the small-scale Gaussian coordinate noise, i.e. recovering $\mathcal{C}_a$ from $\tilde{\mathcal{C}}$, rather than recovering $\mathcal{C}$. Therefore, the addition of torsion augmentations does not alter the fact that 3D denoising clusters closely resembled conformations.
On the other hand, the torsion augmentation not only efficiently provides diverse low-energy conformations for the cross-modal distillation module, but also enhances 3D denoising by enlarging sampling coverage and raising force accuracy, as validated in~\cite{feng2023fractional}.  
Although the combination of the torsion augmentation and traditional denoising results in the hybrid noise strategy in fractional denoising~\cite{feng2023fractional}, our derivation through augmentation is distinct from the previous interpretation of force learning. 

\subsection{Cross-modal Distillation Module}\label{sec: Cross-modal Distillation Module}
To achieve a hierarchical molecular representation, we employ cross-modal contrastive learning, distilling knowledge from the 2D to 3D. Specifically, the encoded 3D conformation representation is denoted as $\boldsymbol{z}^{c} = h_p ( f^{3d} ( \tilde{\mathcal{C}} ))$, where $h_p$ refers to the alignment head used to align dimensions with 2D embeddings. $\tilde{\mathcal{C}}$ is the noisy conformation introduced in the torsion augmented denoising module.
Simultaneously, the encoded 2D molecular graph representation is denoted as $\boldsymbol{z}^{g} = f^{2d} (\mathcal{G})$. The cross-modal contrastive loss is expressed as 
\begin{footnotesize}
\begin{equation}
\begin{aligned}
\mathcal{L}_{cl} = - \mathbb{E}_{\mathcal{G}, \tilde{\mathcal{C}}}[ \log \frac{e^{\text{cos}(\boldsymbol{z}^{g}, \boldsymbol{z}^{c}) / \tau}}{ \sum\limits_{\boldsymbol{z}_j^{c}} e^{\text{cos}(\boldsymbol{z}^{g}, \boldsymbol{z}_j^{c}) / \tau}} +   \log \frac{e^{\text{cos}(\boldsymbol{z}^{c}, \boldsymbol{z}^{g}) / \tau}}{ \sum\limits_{\boldsymbol{z}_j^{g}} e^{\text{cos}(\boldsymbol{z}^{c}, \boldsymbol{z}_j^{g}) / \tau}}]  ,  
\end{aligned}
\end{equation}

\end{footnotesize}
where $\boldsymbol{z}_j^{c}\in \mathbb{N}^c\cup\boldsymbol{z}^{c}$ and $ \boldsymbol{z}_j^{g} \in \mathbb{N}^g\cup\boldsymbol{z}^{g}$, $\mathbb{N}^c$ and $\mathbb{N}^g$ denotes in-batch negative samples of 2D graphs and 3D conformations, $cos(,)$ represents the cosine similarity function, and $\tau$ is the temperature hyperparameter. In different epochs, the same input $(\mathcal{G},\mathcal{C})$ produces diverse torsion-augmented conformations $\mathcal{C}_a$ that are low-energy conformations largely different from each other. Therefore, the contrastive learning clusters dissimilar conformations of the same molecule.

It's important to highlight that conventional approaches primarily leverage 3D data to assist 2D networks in addressing 2D downstream tasks~\cite{stark20223d, liu2021pre}. However, to achieve hierarchical clustering representation, we need to distill 2D knowledge into the 3D network, and not the other way around. This stems from the fact that the most fine-grained hierarchy, distinguishing different equilibriums, cannot be reflected in 2D inputs. This difference reflects the uniqueness of our motivation and methodology.


The ultimate loss, denoted as $\mathcal{L}$, is a sum of the three previously defined losses: $\mathcal{L} = \mathcal{L}_{dn} + \mathcal{L}_{m} + \mathcal{L}_{cl}$. 
 As a result, the learned 3D representation contains knowledge of three distinct hierarchical views of molecules, promising to benefit molecular properties at different scales. Therefore after pre-training, we fine-tune the encoder $f^{3d}$ to adapt it for diverse downstream tasks. Pseudocodes and hyperparameter settings can be found in appendix~\ref{sec:pseodocode} and \ref{app: setting}.

\section{Experiments}
\subsection{Settings}

We use the pre-trained 3D encoder as the backbone for downstream tasks, but employing different heads for different tasks, we employ a simple 2-layer MLP head for MoleculeNet tasks, and use an equivariant head as defined in ~\cite{tholke2022torchmd} for quantum tasks including QM9, MD17 and MD22.

Our hyperparameter selection methods are in line with previous methods. We manually select pre-training and fine-tuning hyperparameters for QM9, MD17, and MD22 to ensure a consistent decrease in pre-training loss and optimal performance on the validation set. For tasks within MoleculeNet, we utilize grid search to identify the most suitable hyperparameters. We provide the selected results and search space in Appendix~\ref{app: setting}.



\subsubsection{Datasets}
Our pre-training dataset consists of 15 million molecules sourced from \citep{nakata2017pubchemqc, zhou2023unimol}. We systematically assess the effectiveness of our model across two different task categories: quantum mechanical tasks, including QM9~\citep{ramakrishnan2014quantum, ruddigkeit2012enumeration}, MD17~\citep{chmiela2017machine} as well as MD22~\citep{Chmiela2023MD22}, and physicochemical and biological property prediction tasks, focusing on the MoleculeNet dataset~\citep{wu2018moleculenet}. Details about downstream datasets can be found in appendix~\ref{sec:app data}. Since we utilize the 3D encoder to fine-tune for downstream tasks, we rely on 3D conformations generated by RDKit for MoleculeNet data, following previous methods~\cite{yu2023unified,zhou2023unimol,fang2022geometry}.

\subsubsection{Baselines}
Our baselines cover three typical categories within molecular pre-training. The first category comprises denoising methods, including Transformer-M~\citep{luo2022one}, SE(3)-DDM~\citep{liu2022molecular}, 3D-EMGP~\citep{jiao2023energy}, Coord~\citep{zaidi2022pre}, Frad~\citep{feng2023fractional}, and Coord(NonEq)~\citep{coordnoneq2023}. The second category consists of multimodal pre-training methods that aim to fusion molecular 2D graphs and 3D conformations, such as 3D InfoMax~\citep{stark20223d}, GraphMVP~\citep{liu2021pre}, MoleculeSDE~\citep{liu2023group}, MoleBLEND~\citep{yu2023unified}, and MoleculeJAE~\citep{du2023molecule}. Notably, 3D InfoMax, GraphMVP, and MoleculeSDE employ contrastive learning between 2D graphs and 3D conformations as a pre-training strategy. The final category encompasses masking methods, which include AttrMask~\citep{hu2020strategies}, GraphMAE~\citep{hou2022graphmae}, GROVER~\citep{rong2020self} and Mole-BERT~\citep{xia2022mole}. Masking methods typically take 2D graphs as input and may not be applicable to quantum downstream tasks that require precise conformation.
\begin{table*}[ht]
 \setlength{\tabcolsep}{0.25em}
\vskip -0.1in
    \caption{Performance (MAE, $\downarrow$) on QM9 quantum tasks. The best results are in bold.}
    \label{table:qm9}
    \vskip 0.15in
    \begin{center}
    \begin{small}
    \scalebox{1.0}{
    \begin{tabular}{llrrrrrrrrrrrr}
    \toprule
     	Methods & Models & $\mu$ (D)	& 	$\alpha$ ($a_0^3$)		&  \makecell[c]{$\epsilon_{HOMO}$ \\(meV)}		& \makecell[c]{$\epsilon_{LUMO}$\\ (meV)}		& \makecell[c]{$\Delta\epsilon$\\ (meV)}	& \makecell[c]{$<R^2>$ \\($a_0^2$)}	& \makecell[c]{ZPVE\\ (meV)	}	& \makecell[c]{$U_0$ \\ (meV)}		& \makecell[c]{$U$ \\ (meV)}		& \makecell[c]{$H$ \\ (meV)}		& \makecell[c]{$G$\\ (meV)} & \makecell[c]{$C_v$\\ ($\frac{cal}{mol K}$)	}
     \\
    \midrule 
    \multirow{5}{*}{\shortstack{Multimodal}} &3D InfoMax & 0.0280 & 0.057 & 25.9 & 21.6 & 42.1& 0.141 & 1.67 & 13.30 & 13.81 & 13.62 & 13.73 & 0.030 \\
    &GraphMVP & 0.0270 &  0.056 & 25.8 & 21.6 & 42.0 & 0.136 & 1.61 & 13.07 & 13.03 & 13.31 & 13.43 & 0.029\\
    &MoleculeSDE & 0.0260  & 0.054  & 25.7 & 21.4 & 41.8 & 0.151 & 1.59 & 12.04 & 12.54 & 12.05 & 13.07 & 0.028 \\
    &MoleculeJAE & 0.0270 & 0.056 & 26.0 & 21.6 & 42.7 & 0.141 & 1.56 & 10.70 & 10.81 & 10.70 & 11.22 & 0.029 \\
    &MoleBLEND &		0.0370 &		0.060	 &	21.5	 &	19.2	 &	34.8	 &	0.417 &		1.58	 &	11.82 &	12.02	& 11.97 &	12.44
    	 &	0.031 \\
    \midrule
        \multirow{4}{*}{\shortstack{3D Denoising}}&Transformer-M &	0.0370 &		0.041 &		17.5 &		16.2 &		27.4 &		0.075 &		\textbf{1.18} &		9.37 &		9.41 &		9.39 &		9.63 &		0.022
     \\
     &SE(3)-DDM 	 &	0.0150	 &	0.046 &		23.5	 &	19.5	 &	40.2	 &	0.122	 &	1.31	 &	6.92 &		6.99	 &	7.09	 &	7.65	 &	0.024
     \\
    &3D-EMGP &	0.0200	 &	0.057	 &	21.3	 &	18.2	 &	37.1	 &	\textbf{0.092}	 &	1.38	 &	8.60 &		8.60	 &	8.70	 &	9.30	 &	0.026
    \\
    &Frad &		0.0100 &		0.037	 &	15.3	 &	13.7	 &	27.8	 &	0.342 &		1.42	 &	5.33 &	5.62	& 5.55 &	6.19
    	 &	0.020 \\
   \midrule
    & UniCorn & \textbf{0.0085} &	\textbf{0.036} &	\textbf{13.0} &	\textbf{11.9} &	\textbf{24.9} &	0.326 &	1.40 &	\textbf{3.99} &	\textbf{3.95} & \textbf{3.94} & \textbf{5.09} &  \textbf{0.019} \\
    
    \bottomrule
    \end{tabular}
    }
    \end{small}
    \end{center}
    \vskip -0.2in
\end{table*}

\begin{table*}[ht]
\setlength{\tabcolsep}{4pt}
\caption{Performance (ROC-AUC \%, $\uparrow$) on MoleculeNet biological classification tasks. The best results are in bold.}
\label{table:molnet}
\vskip 0.15in
\begin{center}
\begin{small}
\scalebox{1.0}{
\begin{tabular}{ll rrrrrrrrrr}
\toprule
Methods & Models & \makecell[c]{BBBP} & \makecell[c]{Tox21} & \makecell[c]{MUV} & \makecell[c]{BACE} &\makecell[c]{ ToxCast} &\makecell[c]{ SIDER} & \makecell[c]{ClinTox} &\makecell[c]{ HIV} & \makecell[c]{Avg.}\\
\midrule
\multirow{4}{*}{\shortstack{Graph\\Masking}} & AttrMask  & 65.0$\pm$2.3 & 74.8$\pm$0.2 & 73.4$\pm$2.0 & 79.7$\pm$0.3 & 62.9$\pm$0.1 & 61.2$\pm$0.1 & 87.7$\pm$1.1 & 76.8$\pm$0.5 & 72.7 \\
&GROVER & 70.0$\pm$0.1 & 74.3$\pm$0.1 &  67.3$\pm$1.8 & 82.6$\pm$0.7 & 65.4$\pm$0.4 & 64.8$\pm$0.6 & 81.2$\pm$3.0 & 62.5$\pm$0.9 & 71.0 \\
& GraphMAE & 72.0$\pm$0.6 & 75.5$\pm$0.6 & 76.3$\pm$2.4 & 83.1$\pm$0.9 & 64.1$\pm$0.3 & 60.3$\pm$1.1 & 82.3$\pm$1.2 & 77.2$\pm$1.0 & 73.9 \\
& Mole-BERT & 71.9$\pm$1.6 & 76.8$\pm$0.5 & 78.6$\pm$1.8 & 80.8$\pm$1.4 & 64.3$\pm$0.2 & 62.8$\pm$1.1 & 78.9$\pm$3.0 & 78.2$\pm$0.8 & 74.0 \\
\midrule
\multirow{4}{*}{\shortstack{Multimodal}}& 3D InfoMax & 69.1$\pm$1.0 & 74.5$\pm$0.7 & 74.4$\pm$2.4  & 79.7$\pm$1.5 & 64.4$\pm$0.8 & 60.6$\pm$0.7 & 79.9$\pm$3.4 & 76.1$\pm$1.3& 72.3 \\
& GraphMVP  & 68.5$\pm$0.2 & 74.5$\pm$0.4 & 75.0$\pm$1.4& 76.8$\pm$1.1 & 62.7$\pm$0.1 & 62.3$\pm$1.6 & 79.0$\pm$2.5 &  74.8$\pm$1.4  & 71.7 \\
& MoleculeSDE & 71.8$\pm$0.7 & 76.8$\pm$0.3 & 80.9$\pm$0.3 & 79.5$\pm$2.1 & 65.0$\pm$0.2 & 60.8$\pm$0.3 & 87.0$\pm$0.5 &  78.8$\pm$0.9& 75.1 \\
& MoleBLEND & 73.0$\pm$0.8 & 77.8$\pm$0.8 &  77.2$\pm$2.3  & 83.7$\pm$1.4 & 66.1$\pm$0.0 & \textbf{64.9$\pm$0.3} & 87.6$\pm$0.7 &79.0$\pm$0.8& 76.2 \\
\midrule
& UniCorn & \textbf{74.2$\pm$1.1} & \textbf{79.3$\pm$0.5} & \textbf{82.6$\pm$1.0} & \textbf{85.8$\pm$1.2} & \textbf{69.4$\pm$1.1} & 64.0$\pm$1.8 & \textbf{92.1$\pm$0.4} & \textbf{79.8$\pm$0.9} & \textbf{78.4}\\
\bottomrule
\end{tabular}
}
\end{small}
\end{center}
\vskip -0.2in
\end{table*}
\begin{table}[h]
    \caption{Performance (MAE, $\downarrow$) on MD17 force prediction tasks (kcal/mol/ $\mathring{\textnormal{A}}$). The best results are in bold. *: SE(3)-DDM employs the Benzene dataset from \citet{chmiela2018towards}, which differs from the version utilized in our work~\citep{chmiela2017machine}.}
    \label{table:md17}
    \vskip 0.15in
    \centering
    \scalebox{1}{
    \begin{scriptsize}
        \setlength{\tabcolsep}{1pt}
    \begin{tabular}{lccccccccc}
    \toprule
      Models& Aspirin	 & 	Benzene	 & 	Ethanol	 & \makecell[c]{	Malonal\\-dehyde}	 & 	\makecell[c]{Naphtha\\-lene}	 & 	\makecell[c]{Salicy\\-lic Acid}		 & Toluene		 & Uracil	\\ 
        \midrule  	 
   MoleculeJAE & 1.289 & 0.345 & 0.365 & 0.613 & 0.498 & 0.712 & 0.480 & 0.463 \\
   MoleculeSDE & 1.112 & 0.304 & 0.282 & 0.520 & 0.455 & 0.725 & 0.515 &  0.447 \\
   \midrule
   SE(3)-DDM* 	 & 0.453	 & 	-	 & 	0.166	 & 	0.288	 & 	0.129	 & 	0.266	 & 	0.122	 & 	0.183\\ 
      Coord	 & 	0.211	  &  	0.169	  &  	0.096	  &  \textbf{	0.139}	  &  	0.053	  &  	0.109	  &  	0.058	  &  	\textbf{0.074}\\
   Frad 	 & 0.209	 & 	0.199	 & 	0.091	 & 	0.142	 & 	0.053	 & 	0.108	 & 0.054	 & 	0.076\\ 
   \midrule
   UniCorn  &		\textbf{0.168} &		\textbf{0.165}	 &	\textbf{0.086}	 &	0.152	 &	\textbf{0.046} &	\textbf{0.098}	& \textbf{0.052} &	0.084\\
    \bottomrule
    \end{tabular}
    \end{scriptsize}
    }
    \vskip -0.2in
\end{table}
\begin{table}[h!]
\setlength{\tabcolsep}{3pt}
    \caption{Performance (RMSE, $\downarrow$) on MoleculeNet physicochemical regression tasks. The best results are in bold.}
    \label{table:molnet2}
    \vskip 0.15in
    \begin{center}
    \begin{footnotesize}
    \scalebox{0.9}{
    \begin{tabular}{lccc}
    \hline
    Models & \makecell[c]{ESOL} & \makecell[c]{FreeSolv} & \makecell[c]{Lipo} \\
    \hline
    AttrMask & 1.112$\pm$0.048 & - & 0.730$\pm$0.004 \\
    GROVER& 0.983$\pm$0.090 & 2.176$\pm$0.052 & 0.817$\pm$0.008 \\
    \midrule
    3D InfoMax  & 0.894$\pm$0.028 & 2.337$\pm$0.227 & 0.695$\pm$0.012 \\
    GraphMVP & 1.029$\pm$0.033 & - & 0.681$\pm$0.010 \\
    MoleBLEND & 0.831$\pm$0.026 & 1.910$\pm$0.163 & 0.638$\pm$0.004 \\
    \midrule
    UniCorn  &\textbf{0.817$\pm$0.034}&\textbf{1.555$\pm$0.075} & \textbf{0.591$\pm$0.016}\\ 
    \hline
    \end{tabular}
    }
    \end{footnotesize}
    \end{center}
    \vskip -0.2in
\end{table}
\vspace{-10pt}
\subsection{Main Experimental Results}\label{sec:main exp}
\subsubsection{Quantum Tasks}
Table~\ref{table:qm9} presents the performance of 12 regression tasks in QM9. Upon comparing various pre-training methods, an evident observation is that the 3D denoising approach significantly outperforms multimodal pre-training methods on average. 
This observation confirms our previous conjecture that the 3D denoising task is more beneficial to quantum tasks. Further supporting this finding is that denoising holds a physical interpretation, equivalent to learning an approximate force field for molecules~\cite{zaidi2022pre}. Notably, our methods outperform existing denoising baselines, achieving the best performance in 10 out of 12 tasks. This achievement can be attributed to the complementary nature of chemical bond information in 2D graph to 3D conformation, enhancing the performance of quantum tasks. This insight is further validated by~\cite{luo2022one, yu2023unified}.

Tables~\ref{table:md17} presents the performance results for force prediction tasks on MD17. UniCorn outperforms existing denoising and multimodal methods, establishing a new state-of-the-art performance on six out of eight molecules. Additionally, we conduct experiments with UniCorn on more challenging force prediction tasks in MD22, as detailed in sections~\ref{md22 exp}, to demonstrate its universality.



\subsubsection{Biological and Physicochemical Tasks}
Table~\ref{table:molnet} illustrates the performance of our method across 8 biological classification tasks in MoleculeNet. Impressively, our approach attains state-of-the-art results in 7 out of the 8 tasks. On average, our method outperforms the second-best method, MoleBLEND, by a substantial margin (78.4 vs 76.2). Furthermore, Table~\ref{table:molnet2} demonstrates that our method achieves the best performance across all 3 physicochemical regression tasks in MoleculeNet.

The improvement observed in our method, as compared to general masking and multimodal methods, can be attributed to the intrinsic connection between macroscopic properties and microscopic quantum properties, which has been thoroughly elucidated by ~\cite{beaini2023towards, sun2022pemp}. The quantum properties of molecules, describing internal electronic motion and atomic nucleus vibrations at the microscopic level, influence the interaction behavior of the molecules with others. These interactions, in turn, affect the physicochemical and biological properties of the molecules. Therefore, the incorporation of denoising task proves instrumental in not only enhancing quantum task performance but also contributing to the improvement in macroscopic properties.

In summary, UniCorn surpasses previous state-of-the-art methods, delivering optimal results across 33 out of 38 molecular tasks that span a wide range of quantum, physicochemical, and biological domains. This superiority not only emphasizes the compatibility of each pre-training strategy but also highlights their complementary nature, thereby contributing to the learning of a universal molecular representation.



\subsection{Ablation Study}

\subsubsection{Loss Study}
To further validate our proposition that distinct pre-training losses exhibit preferences for specific downstream tasks, we conduct two sets of experiments. Firstly, we perform pre-training with and without denoising, followed by fine-tuning on three QM9 tasks. The results are presented in Table~\ref{table:ablation1}, revealing a significant decline in performance without the denoising loss. This underscores the vital role of denoising for quantum chemical tasks. Secondly, we pre-train the model without the masked fragment loss and the cross-modal distillation loss, followed by fine-tuning on 4 biological and 2 physicochemical tasks in MoleculeNet. The results, shown in Table~\ref{table:ablation2}, demonstrate that incorporating fragment masking and cross-modal distillation can boost performance on biological and physicochemical tasks.
\begin{table}[t]
\setlength{\tabcolsep}{4pt}
    \caption{Performance (MAE, $\downarrow$) on QM9. The top results are in bold. }
    \label{table:ablation1}    
    \begin{center}
    \begin{footnotesize}
    \vskip -0.15in
    \scalebox{0.9}{
    \begin{tabular}{lcccc}
    \toprule
    	 QM9 & \makecell[c]{$\epsilon_{LOMO}$ (meV)}		& \makecell[c]{$\epsilon_{HOMO}$ (meV)}		& \makecell[c]{$\Delta\epsilon$(meV)}	\\
    \midrule
    Train from scratch	&16.7  & 17.6 	  &31.3\\
    UniCorn w/o Denoising  & 14.7& 16.8 &	31.0\\ 
     UniCorn &	\textbf{11.9}&	\textbf{13.0}&	\textbf{24.9}	 \\
    \bottomrule
    \end{tabular}
    }
    \end{footnotesize}
    \end{center}
    \vskip -0.2in
\end{table}

\begin{table}[t]
    \caption{Performance (ROC-AUC \%, $\uparrow$ and RMSE, $\downarrow$) on MoleculeNet. The top results are in bold. M\&C denotes fragment masking and cross-modal distillation.}
    \label{table:ablation2}
    \vspace{0.15in}
    \centering
    \scalebox{0.9}{
    \begin{scriptsize}
        \setlength{\tabcolsep}{2pt}
        \begin{tabular}{lcccccc}
            \toprule
            MoleculeNet & \makecell[c]{BBBP$\uparrow$} & \makecell[c]{BACE$\uparrow$} & \makecell[c]{Tox21$\uparrow$} & \makecell[c]{ToxCast$\uparrow$} & \makecell[c]{ESOL$\downarrow$} & \makecell[c]{Lipo$\downarrow$} \\
            \midrule
            Train from scratch & 68.9 (2.4) & 83.6 (1.8) & 76.1 (0.7) & 65.1 (0.1) & 1.083(0.030) & 0.730(0.016) \\
            \makecell[l]{UniCorn w/o M\&C} & 69.3 (0.9) & 81.4 (0.5) & 77.8 (1.4) & 65.8 (0.7) & 0.825(0.036) & 0.623(0.009) \\
            \makecell[l]{UniCorn} & \textbf{74.2 (1.1)} & \textbf{85.8 (1.2)} & \textbf{79.3 (0.5)} & \textbf{69.4 (1.1)} & \textbf{0.817(0.034)} & \textbf{0.591(0.016)} \\
            \bottomrule
        \end{tabular}
    \end{scriptsize}}
    \vspace{-0.2in}
\end{table}

\begin{figure}[t]
\begin{center}
\centerline{\includegraphics[width=\columnwidth]{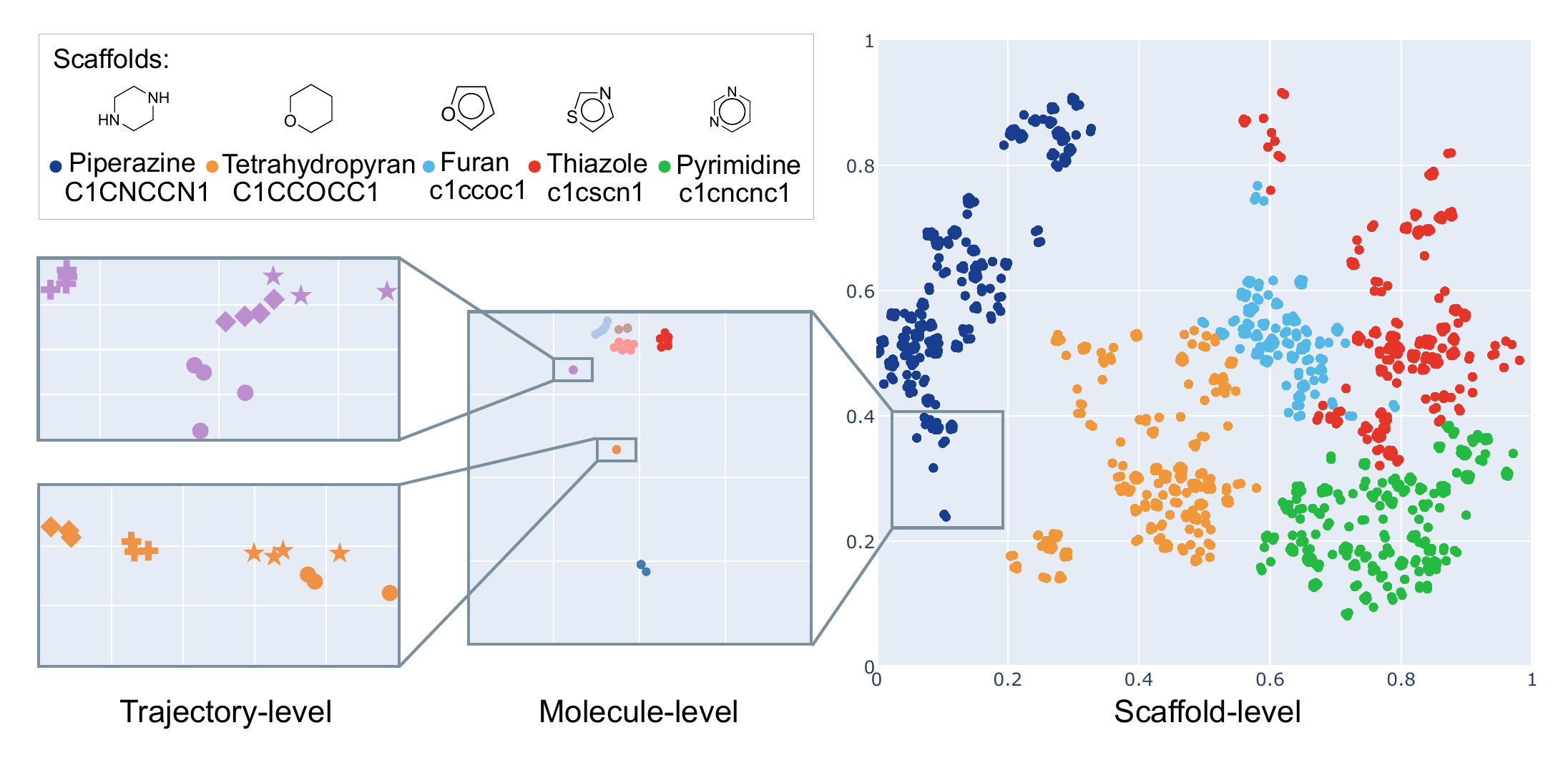}}
\vskip -0.1in
\caption{The visualization showcases hierarchical molecular representations learned by UniCorn, illustrating its ability to achieve effective clustering across different levels.}
\label{fig:abi1}
\end{center}
\vskip -0.5in
\end{figure}

\subsubsection{Hierarchical Feature Visualization}\label{sec:hiervis}

We construct a hierarchical molecular dataset that includes trajectory-level labels representing adjacent similar conformations within the same molecular dynamic trajectory, as well as molecule-level labels denoting multiple trajectories share the same molecule, and scaffold-level labels indicating similar molecules have the same scaffold. A detailed description of the dataset construction process is provided in section~\ref{sec: hierarchical construction}. Subsequently, we employ t-SNE~\citep{van2008visualizing} for visualizing their UniCorn representations.


The clustering results are illustrated in Figure~\ref{fig:abi1}. Moving from left to right, we present the trajectory-level, molecule-level, and scaffold-level clustering results. Given that these levels involve clustering at different resolutions, we have chosen specific areas for a closer examination in the figure. 
At the scaffold level, all five scaffolds are well-separated in general. Notably, heterocycles with similar properties have closer interclass distances, such as non-aromatic rings being situated on the left while aromatic rings are positioned on the right. Zooming into the molecular level, we find the conformations of different molecules are well-clustered. Moreover, when focusing on specific molecules' conformations, we find that similar conformations tend to have closely positioned representations. As a result, UniCorn's molecular representation demonstrates clustering at various levels, affirming the compatibility of the three pre-training modules and their successful realization of multi-view representations.

We also visualize the representations pre-trained by UniCorn and other SSL models across various downstream tasks in section \ref{sec: Downstream  Visualization}. It's important to note that these representations are not fine-tuned, directly reflecting the efficacy of pre-training methods. Once again, the results validate the task-specific preference of existing methods and highlight UniCorn's consistent competence across diverse downstream tasks.



\section{Conclusion}
To address the lack of effective unified models in molecular pre-training, we begin with a theoretical unification of three primary SSL methods: masking, denoising, and contrastive learning, from a contrastive perspective. This comprehension enables us to interpret the three categories of pre-training methods as clustering molecular views at different levels. Subsequently, we expound on how these different levels of views align with specific downstream tasks, thereby elucidating the task-specific preferences of existing methods.
Finally, in pursuit of a universal representation beneficial for diverse downstream tasks, we introduce UniCorn, a novel approach capable of comprehending molecular views at multiple levels. UniCorn demonstrates superior performance compared to existing SSL methods across all three types of tasks. Ablation and visualization analyses further illustrate the compatibility, complementarity, and universality of the multi-view molecular representation learned by UniCorn.

Our work has opened avenues for further exploration in several directions. Firstly, beyond property prediction tasks, whether the UniCorn representation can enhance generation tasks is worth studying.
Secondly, our comprehensive understanding of SSL methods offers a distinctive perspective for elucidating denoising and masking techniques, potentially inspiring advancements in noise addition and masking strategies.
Thirdly, various SSL methods have emerged in other closely related domains~\cite{yang2022scbert,wang2022pre,zhang2022protein,rives2021biological}, yet their understanding and relevance to downstream tasks remain unclear. We aspire that our work can stimulate efforts to establish the connection between pre-training methods and downstream tasks, thereby fostering the development of new foundational models in these domains.

\section*{Acknowledgements}
This work is supported by National Key R\&D Program of China No.2021YFF1201600 and Beijing Academy of Artificial Intelligence (BAAI).

We thank anonymous reviewers for constructive and helpful discussions.

\section*{Impact Statement}
This paper presents work whose goal is to advance the field of Machine Learning. There are many potential societal consequences of our work, none which we feel must be specifically highlighted here.

\bibliography{example_paper}

\begin{thebibliography}{86}
\providecommand{\natexlab}[1]{#1}
\providecommand{\url}[1]{\texttt{#1}}
\expandafter\ifx\csname urlstyle\endcsname\relax
  \providecommand{\doi}[1]{doi: #1}\else
  \providecommand{\doi}{doi: \begingroup \urlstyle{rm}\Url}\fi

\bibitem[Allen et~al.(1996)Allen, Harris, and Taylor]{allen1996conformer}
Allen, F.~H., Harris, S.~E., and Taylor, R.
\newblock Comparison of conformer distributions in the crystalline state with conformational energies calculated by ab initio techniques.
\newblock \emph{Journal of computer-aided molecular design}, 10:\penalty0 247--254, 1996.

\bibitem[Assran et~al.(2023)Assran, Balestriero, Duval, Bordes, Misra, Bojanowski, Vincent, Rabbat, and Ballas]{2022hiddenuniform}
Assran, M., Balestriero, R., Duval, Q., Bordes, F., Misra, I., Bojanowski, P., Vincent, P., Rabbat, M., and Ballas, N.
\newblock The hidden uniform cluster prior in self-supervised learning.
\newblock 2023.

\bibitem[Balestriero et~al.(2023)Balestriero, Ibrahim, Sobal, Morcos, Shekhar, Goldstein, Bordes, Bardes, Mialon, Tian, et~al.]{2023cookbook}
Balestriero, R., Ibrahim, M., Sobal, V., Morcos, A., Shekhar, S., Goldstein, T., Bordes, F., Bardes, A., Mialon, G., Tian, Y., et~al.
\newblock A cookbook of self-supervised learning.
\newblock \emph{arXiv preprint arXiv:2304.12210}, 2023.

\bibitem[Bardes et~al.(2022)Bardes, Ponce, and LeCun]{bardes2021vicreg}
Bardes, A., Ponce, J., and LeCun, Y.
\newblock Vicreg: Variance-invariance-covariance regularization for self-supervised learning.
\newblock 2022.

\bibitem[Beaini et~al.(2023)Beaini, Huang, Cunha, Moisescu-Pareja, Dymov, Maddrell-Mander, McLean, Wenkel, M{\"u}ller, Mohamud, et~al.]{beaini2023towards}
Beaini, D., Huang, S., Cunha, J.~A., Moisescu-Pareja, G., Dymov, O., Maddrell-Mander, S., McLean, C., Wenkel, F., M{\"u}ller, L., Mohamud, J.~H., et~al.
\newblock Towards foundational models for molecular learning on large-scale multi-task datasets.
\newblock \emph{arXiv preprint arXiv:2310.04292}, 2023.

\bibitem[Brameld et~al.(2008)Brameld, Kuhn, Reuter, and Stahl]{brameld2008conformation}
Brameld, K.~A., Kuhn, B., Reuter, D.~C., and Stahl, M.
\newblock Small molecule conformational preferences derived from crystal structure data. a medicinal chemistry focused analysis.
\newblock \emph{Journal of chemical information and modeling}, 48\penalty0 (1):\penalty0 1--24, 2008.

\bibitem[Breccia \& Alimena(2010)Breccia and Alimena]{breccia2010nilotinib}
Breccia, M. and Alimena, G.
\newblock Nilotinib: a second-generation tyrosine kinase inhibitor for chronic myeloid leukemia.
\newblock \emph{Leukemia research}, 34\penalty0 (2):\penalty0 129--134, 2010.

\bibitem[Caron et~al.(2021)Caron, Touvron, Misra, J{\'e}gou, Mairal, Bojanowski, and Joulin]{caron2021dino}
Caron, M., Touvron, H., Misra, I., J{\'e}gou, H., Mairal, J., Bojanowski, P., and Joulin, A.
\newblock Emerging properties in self-supervised vision transformers.
\newblock In \emph{Proceedings of the IEEE/CVF international conference on computer vision}, pp.\  9650--9660, 2021.

\bibitem[Chen et~al.(2020)Chen, Kornblith, Norouzi, and Hinton]{chen2020simclr}
Chen, T., Kornblith, S., Norouzi, M., and Hinton, G.
\newblock A simple framework for contrastive learning of visual representations.
\newblock In \emph{International conference on machine learning}, pp.\  1597--1607. PMLR, 2020.

\bibitem[Chen \& He(2021)Chen and He]{chen2021simsiam}
Chen, X. and He, K.
\newblock Exploring simple siamese representation learning.
\newblock In \emph{Proceedings of the IEEE/CVF conference on computer vision and pattern recognition}, pp.\  15750--15758, 2021.

\bibitem[Chithrananda et~al.(2020)Chithrananda, Grand, and Ramsundar]{chithrananda2020chemberta}
Chithrananda, S., Grand, G., and Ramsundar, B.
\newblock Chemberta: large-scale self-supervised pretraining for molecular property prediction.
\newblock \emph{arXiv preprint arXiv:2010.09885}, 2020.

\bibitem[Chmiela et~al.(2017)Chmiela, Tkatchenko, Sauceda, Poltavsky, Sch{\"u}tt, and M{\"u}ller]{chmiela2017machine}
Chmiela, S., Tkatchenko, A., Sauceda, H.~E., Poltavsky, I., Sch{\"u}tt, K.~T., and M{\"u}ller, K.-R.
\newblock Machine learning of accurate energy-conserving molecular force fields.
\newblock \emph{Science advances}, 3\penalty0 (5):\penalty0 e1603015, 2017.

\bibitem[Chmiela et~al.(2018)Chmiela, Sauceda, M{\"u}ller, and Tkatchenko]{chmiela2018towards}
Chmiela, S., Sauceda, H.~E., M{\"u}ller, K.-R., and Tkatchenko, A.
\newblock Towards exact molecular dynamics simulations with machine-learned force fields.
\newblock \emph{Nature communications}, 9\penalty0 (1):\penalty0 1--10, 2018.

\bibitem[Chmiela et~al.(2023)Chmiela, Vassilev-Galindo, Unke, Kabylda, Sauceda, Tkatchenko, and Müller]{Chmiela2023MD22}
Chmiela, S., Vassilev-Galindo, V., Unke, O.~T., Kabylda, A., Sauceda, H.~E., Tkatchenko, A., and Müller, K.-R.
\newblock Accurate global machine learning force fields for molecules with hundreds of atoms.
\newblock \emph{Science Advances}, 9\penalty0 (2):\penalty0 eadf0873, 2023.
\newblock \doi{10.1126/sciadv.adf0873}.
\newblock URL \url{https://www.science.org/doi/abs/10.1126/sciadv.adf0873}.

\bibitem[Davies \& Bouldin(1979)Davies and Bouldin]{davies1979cluster}
Davies, D.~L. and Bouldin, D.~W.
\newblock A cluster separation measure.
\newblock \emph{IEEE transactions on pattern analysis and machine intelligence}, \penalty0 (2):\penalty0 224--227, 1979.

\bibitem[Dean(1995)]{Dean1995similarity2}
Dean, P.~M.
\newblock \emph{Defining molecular similarity and complementarity for drug design}, pp.\  1--23.
\newblock Springer Netherlands, Dordrecht, 1995.
\newblock ISBN 978-94-011-1350-2.
\newblock \doi{10.1007/978-94-011-1350-2_1}.
\newblock URL \url{https://doi.org/10.1007/978-94-011-1350-2_1}.

\bibitem[Degen et~al.(2008)Degen, Wegscheid-Gerlach, Zaliani, and Rarey]{degen2008art}
Degen, J., Wegscheid-Gerlach, C., Zaliani, A., and Rarey, M.
\newblock On the art of compiling and using'drug-like'chemical fragment spaces.
\newblock \emph{ChemMedChem: Chemistry Enabling Drug Discovery}, 3\penalty0 (10):\penalty0 1503--1507, 2008.

\bibitem[Druker et~al.(2001)Druker, Talpaz, Resta, Peng, Buchdunger, Ford, Lydon, Kantarjian, Capdeville, Ohno-Jones, et~al.]{druker2001imatinib}
Druker, B.~J., Talpaz, M., Resta, D.~J., Peng, B., Buchdunger, E., Ford, J.~M., Lydon, N.~B., Kantarjian, H., Capdeville, R., Ohno-Jones, S., et~al.
\newblock Efficacy and safety of a specific inhibitor of the bcr-abl tyrosine kinase in chronic myeloid leukemia.
\newblock \emph{New England Journal of Medicine}, 344\penalty0 (14):\penalty0 1031--1037, 2001.

\bibitem[Du et~al.(2023)Du, Chen, Zhang, Ma, and Liu]{du2023molecule}
Du, W., Chen, J., Zhang, X., Ma, Z., and Liu, S.
\newblock Molecule joint auto-encoding: Trajectory pretraining with 2d and 3d diffusion.
\newblock \emph{arXiv preprint arXiv:2312.03475}, 2023.

\bibitem[Eastman et~al.(2013)Eastman, Friedrichs, Chodera, Radmer, Bruns, Ku, Beauchamp, Lane, Wang, Shukla, et~al.]{eastman2013openmm}
Eastman, P., Friedrichs, M.~S., Chodera, J.~D., Radmer, R.~J., Bruns, C.~M., Ku, J.~P., Beauchamp, K.~A., Lane, T.~J., Wang, L.-P., Shukla, D., et~al.
\newblock Openmm 4: a reusable, extensible, hardware independent library for high performance molecular simulation.
\newblock \emph{Journal of chemical theory and computation}, 9\penalty0 (1):\penalty0 461--469, 2013.

\bibitem[Erlanson et~al.(2004)Erlanson, McDowell, and O'Brien]{erlanson2004fragment}
Erlanson, D.~A., McDowell, R.~S., and O'Brien, T.
\newblock Fragment-based drug discovery.
\newblock \emph{Journal of medicinal chemistry}, 47\penalty0 (14):\penalty0 3463--3482, 2004.

\bibitem[Erlanson et~al.(2016)Erlanson, Fesik, Hubbard, Jahnke, and Jhoti]{erlanson2016twenty}
Erlanson, D.~A., Fesik, S.~W., Hubbard, R.~E., Jahnke, W., and Jhoti, H.
\newblock Twenty years on: the impact of fragments on drug discovery.
\newblock \emph{Nature reviews Drug discovery}, 15\penalty0 (9):\penalty0 605--619, 2016.

\bibitem[Fang et~al.(2022)Fang, Liu, Lei, He, Zhang, Zhou, Wang, Wu, and Wang]{fang2022geometry}
Fang, X., Liu, L., Lei, J., He, D., Zhang, S., Zhou, J., Wang, F., Wu, H., and Wang, H.
\newblock Geometry-enhanced molecular representation learning for property prediction.
\newblock \emph{Nature Machine Intelligence}, 4\penalty0 (2):\penalty0 127--134, 2022.

\bibitem[Feng et~al.(2023{\natexlab{a}})Feng, Ni, Lan, Ma, and Ma]{feng2023fractional}
Feng, S., Ni, Y., Lan, Y., Ma, Z.-M., and Ma, W.-Y.
\newblock Fractional denoising for 3{D} molecular pre-training.
\newblock In \emph{International Conference on Machine Learning}, pp.\  9938--9961. PMLR, 2023{\natexlab{a}}.

\bibitem[Feng et~al.(2023{\natexlab{b}})Feng, Ni, Lan, Ma, and Ma]{pmlr-v202-feng23c}
Feng, S., Ni, Y., Lan, Y., Ma, Z.-M., and Ma, W.-Y.
\newblock Fractional denoising for 3{D} molecular pre-training.
\newblock In \emph{Proceedings of the 40th International Conference on Machine Learning}, volume 202 of \emph{Proceedings of Machine Learning Research}, pp.\  9938--9961. PMLR, 23--29 Jul 2023{\natexlab{b}}.
\newblock URL \url{https://proceedings.mlr.press/v202/feng23c.html}.

\bibitem[Feng et~al.(2023{\natexlab{c}})Feng, Yang, Ma, and Lan]{feng2023unimap}
Feng, S., Yang, L., Ma, W., and Lan, Y.
\newblock Unimap: Universal smiles-graph representation learning.
\newblock \emph{arXiv preprint arXiv:2310.14216}, 2023{\natexlab{c}}.

\bibitem[Galanti et~al.(2022{\natexlab{a}})Galanti, Gy{\"o}rgy, and Hutter]{galanti2021fewshot}
Galanti, T., Gy{\"o}rgy, A., and Hutter, M.
\newblock On the role of neural collapse in transfer learning.
\newblock 2022{\natexlab{a}}.

\bibitem[Galanti et~al.(2022{\natexlab{b}})Galanti, Gy{\"o}rgy, and Hutter]{galanti2022transfer}
Galanti, T., Gy{\"o}rgy, A., and Hutter, M.
\newblock Improved generalization bounds for transfer learning via neural collapse.
\newblock In \emph{First Workshop on Pre-training: Perspectives, Pitfalls, and Paths Forward at ICML 2022}, 2022{\natexlab{b}}.

\bibitem[Gao et~al.(2022)Gao, Tan, Wu, and Li]{gao2022cosp}
Gao, Z., Tan, C., Wu, L., and Li, S.~Z.
\newblock Cosp: Co-supervised pretraining of pocket and ligand.
\newblock \emph{arXiv preprint arXiv:2206.12241}, 2022.

\bibitem[Garrido et~al.(2023)Garrido, Chen, Bardes, Najman, and Lecun]{garrido2022duality}
Garrido, Q., Chen, Y., Bardes, A., Najman, L., and Lecun, Y.
\newblock On the duality between contrastive and non-contrastive self-supervised learning.
\newblock 2023.

\bibitem[Godwin et~al.(2021)Godwin, Schaarschmidt, Gaunt, Sanchez-Gonzalez, Rubanova, Velivckovi'c, Kirkpatrick, and Battaglia]{godwin2021simple}
Godwin, J., Schaarschmidt, M., Gaunt, A., Sanchez-Gonzalez, A., Rubanova, Y., Velivckovi'c, P., Kirkpatrick, J., and Battaglia, P.~W.
\newblock Simple gnn regularisation for 3{D} molecular property prediction and beyond.
\newblock In \emph{International Conference on Learning Representations}, 2021.
\newblock URL \url{https://api.semanticscholar.org/CorpusID:247450503}.

\bibitem[Grill et~al.(2020)Grill, Strub, Altch{\'e}, Tallec, Richemond, Buchatskaya, Doersch, Avila~Pires, Guo, Gheshlaghi~Azar, et~al.]{grill2020byol}
Grill, J.-B., Strub, F., Altch{\'e}, F., Tallec, C., Richemond, P., Buchatskaya, E., Doersch, C., Avila~Pires, B., Guo, Z., Gheshlaghi~Azar, M., et~al.
\newblock Bootstrap your own latent-a new approach to self-supervised learning.
\newblock \emph{Advances in neural information processing systems}, 33:\penalty0 21271--21284, 2020.

\bibitem[HaoChen et~al.(2021)HaoChen, Wei, Gaidon, and Ma]{haochen2021spectralcl}
HaoChen, J.~Z., Wei, C., Gaidon, A., and Ma, T.
\newblock Provable guarantees for self-supervised deep learning with spectral contrastive loss.
\newblock \emph{Advances in Neural Information Processing Systems}, 34:\penalty0 5000--5011, 2021.

\bibitem[Hou et~al.(2022)Hou, Liu, Cen, Dong, Yang, Wang, and Tang]{hou2022graphmae}
Hou, Z., Liu, X., Cen, Y., Dong, Y., Yang, H., Wang, C., and Tang, J.
\newblock Graphmae: Self-supervised masked graph autoencoders.
\newblock In \emph{Proceedings of the 28th ACM SIGKDD Conference on Knowledge Discovery and Data Mining}, pp.\  594--604, 2022.

\bibitem[Hu et~al.(2020)Hu, Liu, Gomes, Zitnik, Liang, Pande, and Leskovec]{hu2020strategies}
Hu, W., Liu, B., Gomes, J., Zitnik, M., Liang, P., Pande, V., and Leskovec, J.
\newblock Strategies for pre-training graph neural networks.
\newblock In \emph{International Conference on Learning Representations (ICLR)}, 2020.

\bibitem[Huang et~al.(2023)Huang, Yi, Zhao, and Jiang]{yimingyang2021}
Huang, W., Yi, M., Zhao, X., and Jiang, Z.
\newblock Towards the generalization of contrastive self-supervised learning.
\newblock 2023.

\bibitem[Jiao et~al.(2023)Jiao, Han, Huang, Rong, and Liu]{jiao2023energy}
Jiao, R., Han, J., Huang, W., Rong, Y., and Liu, Y.
\newblock Energy-motivated equivariant pretraining for 3d molecular graphs.
\newblock In \emph{Proceedings of the AAAI Conference on Artificial Intelligence}, volume~37, pp.\  8096--8104, 2023.

\bibitem[Johnson et~al.(1990)Johnson, Maggiora, et~al.]{johnson1990similarity1}
Johnson, M.~A., Maggiora, G.~M., et~al.
\newblock Concepts and applications of molecular similarity.
\newblock 1990.
\newblock URL \url{https://cir.nii.ac.jp/crid/1130000796218783104}.

\bibitem[Kantarjian et~al.(2021)Kantarjian, Hughes, Larson, Kim, Issaragrisil, le~Coutre, Etienne, Boquimpani, Pasquini, Clark, et~al.]{kantarjian2021long}
Kantarjian, H.~M., Hughes, T.~P., Larson, R.~A., Kim, D.-W., Issaragrisil, S., le~Coutre, P., Etienne, G., Boquimpani, C., Pasquini, R., Clark, R.~E., et~al.
\newblock Long-term outcomes with frontline nilotinib versus imatinib in newly diagnosed chronic myeloid leukemia in chronic phase: Enestnd 10-year analysis.
\newblock \emph{Leukemia}, 35\penalty0 (2):\penalty0 440--453, 2021.

\bibitem[Kauzmann(2013)]{kauzmann2013quantumbook}
Kauzmann, W.
\newblock \emph{Quantum chemistry: an introduction}.
\newblock Elsevier, 2013.

\bibitem[Kirillov et~al.(2023)Kirillov, Mintun, Ravi, Mao, Rolland, Gustafson, Xiao, Whitehead, Berg, Lo, Doll{\'a}r, and Girshick]{kirillov2023segany}
Kirillov, A., Mintun, E., Ravi, N., Mao, H., Rolland, C., Gustafson, L., Xiao, T., Whitehead, S., Berg, A.~C., Lo, W.-Y., Doll{\'a}r, P., and Girshick, R.
\newblock Segment anything.
\newblock \emph{arXiv:2304.02643}, 2023.

\bibitem[LANDAU \& LIFSHITZ(1975)LANDAU and LIFSHITZ]{LANDAU197589}
LANDAU, L. and LIFSHITZ, E.
\newblock Chapter 5 - constant electromagnetic fields.
\newblock In LANDAU, L. and LIFSHITZ, E. (eds.), \emph{The Classical Theory of Fields (Fourth Edition)}, volume~2 of \emph{Course of Theoretical Physics}, pp.\  89--108. Pergamon, Amsterdam, fourth edition edition, 1975.
\newblock ISBN 978-0-08-025072-4.
\newblock \doi{https://doi.org/10.1016/B978-0-08-025072-4.50012-5}.
\newblock URL \url{https://www.sciencedirect.com/science/article/pii/B9780080250724500125}.

\bibitem[Le~Questel(2023)]{le2023conformations}
Le~Questel, J.-Y.
\newblock Conformations and physicochemical properties of biological ligands in various environments, 2023.

\bibitem[Li et~al.(2023{\natexlab{a}})Li, Gan, Yang, Yang, Li, Wang, and Gao]{li2023multimodal}
Li, C., Gan, Z., Yang, Z., Yang, J., Li, L., Wang, L., and Gao, J.
\newblock Multimodal foundation models: From specialists to general-purpose assistants.
\newblock \emph{arXiv preprint arXiv:2309.10020}, 1\penalty0 (2):\penalty0 2, 2023{\natexlab{a}}.

\bibitem[Li et~al.(2023{\natexlab{b}})Li, Wu, Sun, Chen, Tian, Zhu, Meng, Zheng, and Wang]{maskgraph}
Li, J., Wu, R., Sun, W., Chen, L., Tian, S., Zhu, L., Meng, C., Zheng, Z., and Wang, W.
\newblock What's behind the mask: Understanding masked graph modeling for graph autoencoders.
\newblock In \emph{Proceedings of the 29th ACM SIGKDD Conference on Knowledge Discovery and Data Mining}, KDD '23, pp.\  1268–1279, New York, NY, USA, 2023{\natexlab{b}}. Association for Computing Machinery.
\newblock ISBN 9798400701030.
\newblock \doi{10.1145/3580305.3599546}.
\newblock URL \url{https://doi.org/10.1145/3580305.3599546}.

\bibitem[Li et~al.(2022)Li, Zhou, Xu, Dou, and Xiong]{li2022geomgcl}
Li, S., Zhou, J., Xu, T., Dou, D., and Xiong, H.
\newblock Geomgcl: Geometric graph contrastive learning for molecular property prediction.
\newblock In \emph{Proceedings of the AAAI conference on artificial intelligence}, volume~36, pp.\  4541--4549, 2022.

\bibitem[Liu et~al.(2021)Liu, Wang, Liu, Lasenby, Guo, and Tang]{liu2021pre}
Liu, S., Wang, H., Liu, W., Lasenby, J., Guo, H., and Tang, J.
\newblock Pre-training molecular graph representation with 3{D} geometry.
\newblock In \emph{International Conference on Learning Representations}, 2021.

\bibitem[Liu et~al.(2022)Liu, Guo, and Tang]{liu2022molecular}
Liu, S., Guo, H., and Tang, J.
\newblock Molecular geometry pretraining with se (3)-invariant denoising distance matching.
\newblock In \emph{International Conference on Learning Representations}, 2022.

\bibitem[Liu et~al.(2023{\natexlab{a}})Liu, Du, Ma, Guo, and Tang]{liu2023group}
Liu, S., Du, W., Ma, Z.-M., Guo, H., and Tang, J.
\newblock A group symmetric stochastic differential equation model for molecule multi-modal pretraining.
\newblock In \emph{International Conference on Machine Learning}, pp.\  21497--21526. PMLR, 2023{\natexlab{a}}.

\bibitem[Liu et~al.(2023{\natexlab{b}})Liu, Nie, Wang, Lu, Qiao, Liu, Tang, Xiao, and Anandkumar]{liu2023multi}
Liu, S., Nie, W., Wang, C., Lu, J., Qiao, Z., Liu, L., Tang, J., Xiao, C., and Anandkumar, A.
\newblock Multi-modal molecule structure--text model for text-based retrieval and editing.
\newblock \emph{Nature Machine Intelligence}, 5\penalty0 (12):\penalty0 1447--1457, 2023{\natexlab{b}}.

\bibitem[Lu et~al.(2022)Lu, Clark, Zellers, Mottaghi, and Kembhavi]{lu2022unified}
Lu, J., Clark, C., Zellers, R., Mottaghi, R., and Kembhavi, A.
\newblock Unified-io: A unified model for vision, language, and multi-modal tasks.
\newblock \emph{arXiv preprint arXiv:2206.08916}, 2022.

\bibitem[Luo et~al.(2023)Luo, Chen, Xu, Zheng, Liu, Wang, and He]{luo2022one}
Luo, S., Chen, T., Xu, Y., Zheng, S., Liu, T.-Y., Wang, L., and He, D.
\newblock One transformer can understand both 2{D} \& 3{D} molecular data.
\newblock In \emph{The Eleventh International Conference on Learning Representations}, 2023.
\newblock URL \url{https://openreview.net/forum?id=vZTp1oPV3PC}.

\bibitem[Manley et~al.(2005)Manley, Cowan-Jacob, and Mestan]{manley2005advances}
Manley, P.~W., Cowan-Jacob, S.~W., and Mestan, J.
\newblock Advances in the structural biology, design and clinical development of bcr-abl kinase inhibitors for the treatment of chronic myeloid leukaemia.
\newblock \emph{Biochimica et Biophysica Acta (BBA)-Proteins and Proteomics}, 1754\penalty0 (1-2):\penalty0 3--13, 2005.

\bibitem[Murray \& Rees(2009)Murray and Rees]{murray2009rise}
Murray, C.~W. and Rees, D.~C.
\newblock The rise of fragment-based drug discovery.
\newblock \emph{Nature chemistry}, 1\penalty0 (3):\penalty0 187--192, 2009.

\bibitem[Nakata \& Shimazaki(2017)Nakata and Shimazaki]{nakata2017pubchemqc}
Nakata, M. and Shimazaki, T.
\newblock Pubchemqc project: a large-scale first-principles electronic structure database for data-driven chemistry.
\newblock \emph{Journal of chemical information and modeling}, 57\penalty0 (6):\penalty0 1300--1308, 2017.

\bibitem[Ni et~al.(2024)Ni, Feng, Ma, Ma, and Lan]{ni2024sliced}
Ni, Y., Feng, S., Ma, W.-Y., Ma, Z.-M., and Lan, Y.
\newblock Sliced denoising: A physics-informed molecular pre-training method.
\newblock In \emph{The Twelfth International Conference on Learning Representations}, 2024.
\newblock URL \url{https://openreview.net/forum?id=liKkG1zcWq}.

\bibitem[Oord et~al.(2018)Oord, Li, and Vinyals]{oord2018infonce}
Oord, A. v.~d., Li, Y., and Vinyals, O.
\newblock Representation learning with contrastive predictive coding.
\newblock \emph{arXiv preprint arXiv:1807.03748}, 2018.

\bibitem[OpenAI(2022)]{chatgpt}
OpenAI.
\newblock Chat{GPT}.
\newblock 2022.
\newblock URL \url{https://openai.com/blog/chatgpt/}.

\bibitem[OpenAI(2023)]{gpt4}
OpenAI.
\newblock {GPT}-4 technical report.
\newblock 2023.
\newblock URL \url{https://arxiv.org/abs/2303.08774}.

\bibitem[Qi et~al.(2023)Qi, Dong, Fan, Ge, Zhang, Ma, and Yi]{qi2023recon}
Qi, Z., Dong, R., Fan, G., Ge, Z., Zhang, X., Ma, K., and Yi, L.
\newblock Contrast with reconstruct: Contrastive 3{D} representation learning guided by generative pretraining.
\newblock \emph{arXiv preprint arXiv:2302.02318}, 2023.

\bibitem[Ramakrishnan et~al.(2014)Ramakrishnan, Dral, Rupp, and Von~Lilienfeld]{ramakrishnan2014quantum}
Ramakrishnan, R., Dral, P.~O., Rupp, M., and Von~Lilienfeld, O.~A.
\newblock Quantum chemistry structures and properties of 134 kilo molecules.
\newblock \emph{Scientific data}, 1\penalty0 (1):\penalty0 1--7, 2014.

\bibitem[Rives et~al.(2021)Rives, Meier, Sercu, Goyal, Lin, Liu, Guo, Ott, Zitnick, Ma, et~al.]{rives2021biological}
Rives, A., Meier, J., Sercu, T., Goyal, S., Lin, Z., Liu, J., Guo, D., Ott, M., Zitnick, C.~L., Ma, J., et~al.
\newblock Biological structure and function emerge from scaling unsupervised learning to 250 million protein sequences.
\newblock \emph{Proceedings of the National Academy of Sciences}, 118\penalty0 (15):\penalty0 e2016239118, 2021.

\bibitem[Rong et~al.(2020)Rong, Bian, Xu, Xie, Wei, Huang, and Huang]{rong2020self}
Rong, Y., Bian, Y., Xu, T., Xie, W., Wei, Y., Huang, W., and Huang, J.
\newblock Self-supervised graph transformer on large-scale molecular data.
\newblock \emph{Advances in Neural Information Processing Systems}, 33:\penalty0 12559--12571, 2020.

\bibitem[Ruddigkeit et~al.(2012)Ruddigkeit, Van~Deursen, Blum, and Reymond]{ruddigkeit2012enumeration}
Ruddigkeit, L., Van~Deursen, R., Blum, L.~C., and Reymond, J.-L.
\newblock Enumeration of 166 billion organic small molecules in the chemical universe database gdb-17.
\newblock \emph{Journal of chemical information and modeling}, 52\penalty0 (11):\penalty0 2864--2875, 2012.

\bibitem[St{\"a}rk et~al.(2022)St{\"a}rk, Beaini, Corso, Tossou, Dallago, G{\"u}nnemann, and Li{\`o}]{stark20223d}
St{\"a}rk, H., Beaini, D., Corso, G., Tossou, P., Dallago, C., G{\"u}nnemann, S., and Li{\`o}, P.
\newblock 3{D} infomax improves gnns for molecular property prediction.
\newblock In \emph{International Conference on Machine Learning}, pp.\  20479--20502. PMLR, 2022.

\bibitem[Sun et~al.(2022)Sun, Chen, Ma, Huang, Liu, Ma, Ma, and Lan]{sun2022pemp}
Sun, Y., Chen, Y., Ma, W., Huang, W., Liu, K., Ma, Z., Ma, W.-Y., and Lan, Y.
\newblock Pemp: Leveraging physics properties to enhance molecular property prediction.
\newblock In \emph{Proceedings of the 31st ACM International Conference on Information \& Knowledge Management}, pp.\  3505--3513, 2022.

\bibitem[Tao et~al.(2022)Tao, Wang, Zhu, Dong, Song, Huang, and Dai]{tao2022equi1}
Tao, C., Wang, H., Zhu, X., Dong, J., Song, S., Huang, G., and Dai, J.
\newblock Exploring the equivalence of siamese self-supervised learning via a unified gradient framework.
\newblock In \emph{Proceedings of the IEEE/CVF Conference on Computer Vision and Pattern Recognition}, pp.\  14431--14440, 2022.

\bibitem[Th{\"o}lke \& De~Fabritiis(2022)Th{\"o}lke and De~Fabritiis]{tholke2022torchmd}
Th{\"o}lke, P. and De~Fabritiis, G.
\newblock Torchmd-net: equivariant transformers for neural network based molecular potentials.
\newblock \emph{arXiv preprint arXiv:2202.02541}, 2022.

\bibitem[Tian et~al.(2020)Tian, Sun, Poole, Krishnan, Schmid, and Isola]{tian2020makes}
Tian, Y., Sun, C., Poole, B., Krishnan, D., Schmid, C., and Isola, P.
\newblock What makes for good views for contrastive learning?
\newblock \emph{Advances in neural information processing systems}, 33:\penalty0 6827--6839, 2020.

\bibitem[Van~der Maaten \& Hinton(2008)Van~der Maaten and Hinton]{van2008visualizing}
Van~der Maaten, L. and Hinton, G.
\newblock Visualizing data using t-sne.
\newblock \emph{Journal of machine learning research}, 9\penalty0 (11), 2008.

\bibitem[Wang et~al.(2022{\natexlab{a}})Wang, Fei, and Zhou]{wang2022pre}
Wang, D., Fei, Y., and Zhou, H.
\newblock On pre-training language model for antibody.
\newblock In \emph{The Eleventh International Conference on Learning Representations}, 2022{\natexlab{a}}.

\bibitem[Wang et~al.(2019)Wang, Guo, Wang, Sun, and Huang]{wang2019smiles}
Wang, S., Guo, Y., Wang, Y., Sun, H., and Huang, J.
\newblock Smiles-bert: large scale unsupervised pre-training for molecular property prediction.
\newblock In \emph{Proceedings of the 10th ACM international conference on bioinformatics, computational biology and health informatics}, pp.\  429--436, 2019.

\bibitem[Wang et~al.(2022{\natexlab{b}})Wang, Zhang, Wang, Yang, and Lin]{wang2022ladder}
Wang, Y., Zhang, Q., Wang, Y., Yang, J., and Lin, Z.
\newblock Chaos is a ladder: A new understanding of contrastive learning.
\newblock 2022{\natexlab{b}}.
\newblock URL \url{https://openreview.net/forum?id=ECvgmYVyeUz}.

\bibitem[Wang et~al.(2023)Wang, Xu, Li, and Barati~Farimani]{coordnoneq2023}
Wang, Y., Xu, C., Li, Z., and Barati~Farimani, A.
\newblock Denoise pretraining on nonequilibrium molecules for accurate and transferable neural potentials.
\newblock \emph{Journal of Chemical Theory and Computation}, 19\penalty0 (15):\penalty0 5077--5087, 2023.
\newblock \doi{10.1021/acs.jctc.3c00289}.
\newblock URL \url{https://doi.org/10.1021/acs.jctc.3c00289}.
\newblock PMID: 37390120.

\bibitem[Willett et~al.(1998)Willett, Barnard, and Downs]{willett1998similarity3}
Willett, P., Barnard, J.~M., and Downs, G.~M.
\newblock Chemical similarity searching.
\newblock \emph{Journal of chemical information and computer sciences}, 38\penalty0 (6):\penalty0 983--996, 1998.

\bibitem[Wu et~al.(2018)Wu, Ramsundar, Feinberg, Gomes, Geniesse, Pappu, Leswing, and Pande]{wu2018moleculenet}
Wu, Z., Ramsundar, B., Feinberg, E.~N., Gomes, J., Geniesse, C., Pappu, A.~S., Leswing, K., and Pande, V.
\newblock Moleculenet: a benchmark for molecular machine learning.
\newblock \emph{Chemical science}, 9\penalty0 (2):\penalty0 513--530, 2018.

\bibitem[Xia et~al.(2022)Xia, Zhao, Hu, Gao, Tan, Liu, Li, and Li]{xia2022mole}
Xia, J., Zhao, C., Hu, B., Gao, Z., Tan, C., Liu, Y., Li, S., and Li, S.~Z.
\newblock Mole-bert: Rethinking pre-training graph neural networks for molecules.
\newblock In \emph{International Conference on Learning Representations}, 2022.

\bibitem[Yang et~al.(2022)Yang, Wang, Wang, Fang, Tang, Huang, Lu, and Yao]{yang2022scbert}
Yang, F., Wang, W., Wang, F., Fang, Y., Tang, D., Huang, J., Lu, H., and Yao, J.
\newblock scbert as a large-scale pretrained deep language model for cell type annotation of single-cell rna-seq data.
\newblock \emph{Nature Machine Intelligence}, 4\penalty0 (10):\penalty0 852--866, 2022.

\bibitem[Ying et~al.(2021)Ying, Cai, Luo, Zheng, Ke, He, Shen, and Liu]{ying2021transformers}
Ying, C., Cai, T., Luo, S., Zheng, S., Ke, G., He, D., Shen, Y., and Liu, T.-Y.
\newblock Do transformers really perform badly for graph representation?
\newblock \emph{Advances in Neural Information Processing Systems}, 34:\penalty0 28877--28888, 2021.

\bibitem[Yu et~al.(2023)Yu, Zhang, Ni, Feng, Lan, Zhou, and Liu]{yu2023unified}
Yu, Q., Zhang, Y., Ni, Y., Feng, S., Lan, Y., Zhou, H., and Liu, J.
\newblock Unified molecular modeling via modality blending.
\newblock \emph{arXiv preprint arXiv:2307.06235}, 2023.

\bibitem[Zaidi et~al.(2022)Zaidi, Schaarschmidt, Martens, Kim, Teh, Sanchez-Gonzalez, Battaglia, Pascanu, and Godwin]{zaidi2022pre}
Zaidi, S., Schaarschmidt, M., Martens, J., Kim, H., Teh, Y.~W., Sanchez-Gonzalez, A., Battaglia, P., Pascanu, R., and Godwin, J.
\newblock Pre-training via denoising for molecular property prediction.
\newblock In \emph{International Conference on Learning Representations}, 2022.

\bibitem[Zhang et~al.(2022{\natexlab{a}})Zhang, Wang, and Wang]{zhang2022theoryMAE}
Zhang, Q., Wang, Y., and Wang, Y.
\newblock How mask matters: Towards theoretical understandings of masked autoencoders.
\newblock \emph{Advances in Neural Information Processing Systems}, 35:\penalty0 27127--27139, 2022{\natexlab{a}}.

\bibitem[Zhang et~al.(2021)Zhang, Wu, Yang, Wu, Yi, Hsieh, Hou, and Cao]{zhang2021mg}
Zhang, X.-C., Wu, C.-K., Yang, Z.-J., Wu, Z.-X., Yi, J.-C., Hsieh, C.-Y., Hou, T.-J., and Cao, D.-S.
\newblock Mg-bert: leveraging unsupervised atomic representation learning for molecular property prediction.
\newblock \emph{Briefings in bioinformatics}, 22\penalty0 (6):\penalty0 bbab152, 2021.

\bibitem[Zhang et~al.(2022{\natexlab{b}})Zhang, Xu, Jamasb, Chenthamarakshan, Lozano, Das, and Tang]{zhang2022protein}
Zhang, Z., Xu, M., Jamasb, A.~R., Chenthamarakshan, V., Lozano, A., Das, P., and Tang, J.
\newblock Protein representation learning by geometric structure pretraining.
\newblock In \emph{The Eleventh International Conference on Learning Representations}, 2022{\natexlab{b}}.

\bibitem[Zhou et~al.(2023)Zhou, Gao, Ding, Zheng, Xu, Wei, Zhang, and Ke]{zhou2023unimol}
Zhou, G., Gao, Z., Ding, Q., Zheng, H., Xu, H., Wei, Z., Zhang, L., and Ke, G.
\newblock Uni-mol: A universal 3{D} molecular representation learning framework.
\newblock In \emph{The Eleventh International Conference on Learning Representations}, 2023.
\newblock URL \url{https://openreview.net/forum?id=6K2RM6wVqKu}.

\bibitem[Zorich(2015)]{Zorich2015I}
Zorich, V.~A.
\newblock \emph{The Differential Calculus of Functions of Several Variables}, pp.\  427--543.
\newblock Springer Berlin Heidelberg, Berlin, Heidelberg, 2015.
\newblock ISBN 978-3-662-48792-1.
\newblock \doi{10.1007/978-3-662-48792-1_8}.
\newblock URL \url{https://doi.org/10.1007/978-3-662-48792-1_8}.

\end{thebibliography}
\bibliographystyle{icml2024}

\newpage
\appendix
\onecolumn

\section{More Experimental Results}
\subsection{MD22}\label{md22 exp}

In Table~\ref{table:md22}, we present the performance results for force prediction tasks on MD22. UniCorn consistently outperforms the best results across all seven tasks, which include larger molecules with more complex structures, highlighting the broad application and generalization capabilities of our approach.

\begin{table}[h]
\centering
\footnotesize
\setlength{\tabcolsep}{9pt}
    \caption{Performance (MAE, $\downarrow$) on MD22 force prediction tasks(kcal/mol/ $\mathring{\textnormal{A}}$). The best results are in bold.}
    \label{table:md22}
    \begin{tabular}{lccccccc}
    \toprule
      Models &  \makecell[c]{ Ac-Ala3-NHMe}	 & 	\makecell[c]{ DHA}	 & \makecell[c]{ 	AT-AT}	 & \makecell[c]{ Staychose}	 & 	 \makecell[c]{ AT-AT -CG-CG}	 & 	\makecell[c]{ Buckyball}		 &  \makecell[c]{ Double -walled}		 	\\ 
        \hline
      \makecell[l]{Coord(NonEq)}   & 0.102 & 0.135 & 0.288 & 0.673 & 0.657 & 1.751 & 2.515 \\ 
   Frad & 0.073 & 0.078 & 0.233 & 0.231 & 0.308 & 0.430 & 0.729\\
   UniCorn & \textbf{0.056} & \textbf{0.051} & \textbf{0.134} & \textbf{0.127} & \textbf{0.181} & \textbf{0.249} & \textbf{0.473} \\
    \bottomrule
    \end{tabular}
    \vskip -0.2in
\end{table}
\vspace{10pt}

\subsection{Downstream Tasks Feature Visualization}\label{sec: Downstream  Visualization}

\begin{figure*}[h]
\begin{center}
\centerline{\includegraphics[width=\textwidth]{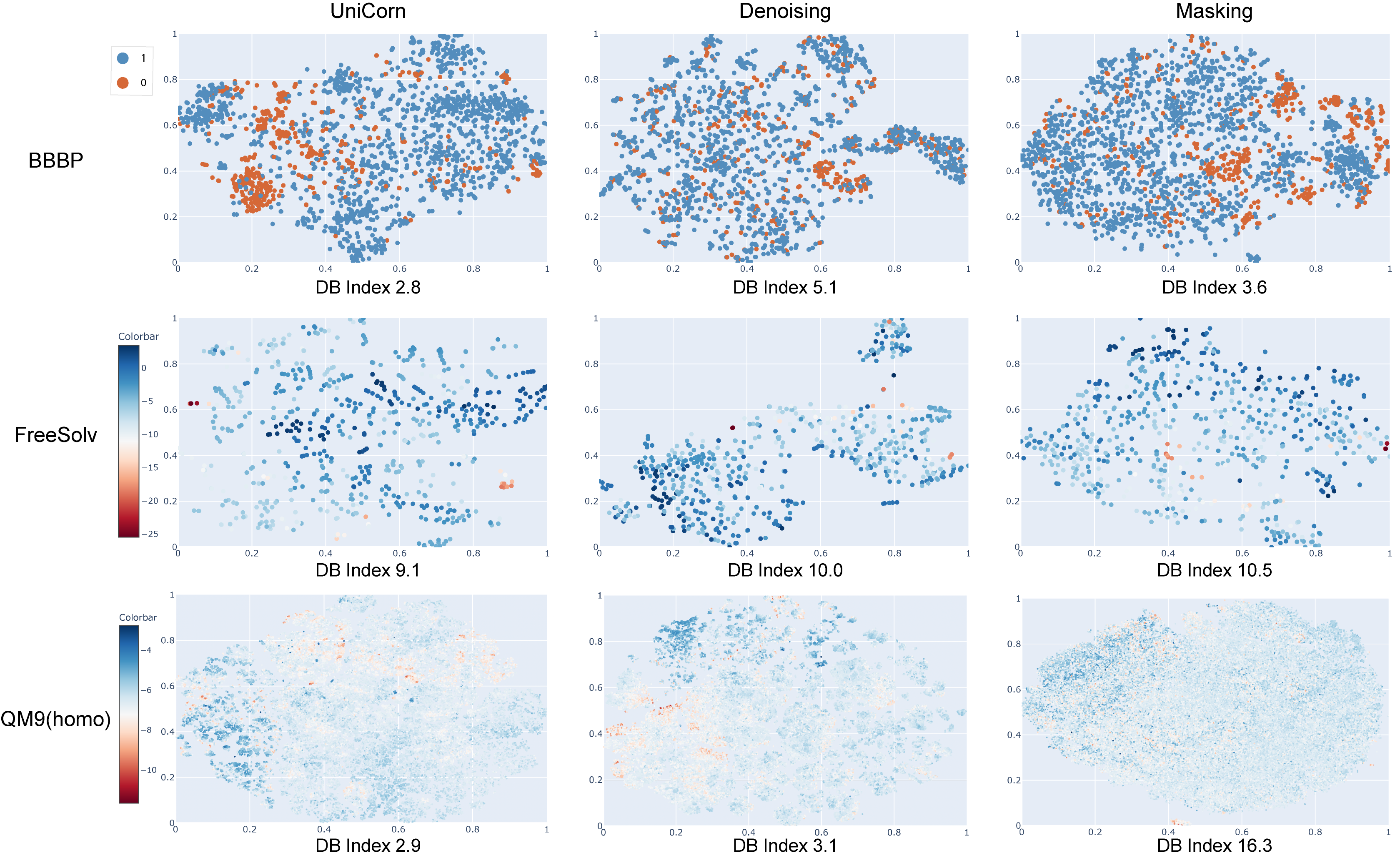}}
\caption{The clustering results of \textbf{unfine-tuned} molecular representations by three distinct methods, across three diverse tasks:  BBBP (biological task), Freesolv (physicochemical task), and homo from QM9 (quantum task). The color indicates the labels of the downstream tasks—discrete binary labels for the BBBP task and continuous labels for Freesolv and QM9. Below each subfigure we present Davies–Bouldin Index to evaluate the performance of clustering results(smaller is better). While the Masking and Denoising methods exhibit a preference for biological and quantum tasks respectively, UniCorn demonstrates the capability to achieve significant clustering results across all types of tasks.}
\label{fig:abi2}
\end{center}
\vskip -10pt
\end{figure*}


To illustrate the correlation between different pre-training methods and various types of downstream tasks analyzed in the previous section~\ref{sec:relation between method&task}, we employed t-SNE to visualize the unfine-tuned features pre-trained by UniCorn, and other two representative methods: Coord~\citep{zaidi2022pre} representing for denoising method, and AttrMask~\citep{hu2020strategies} representing for masking approach, on three typical downstream tasks: the biological task BBBP, the physical chemistry task Freesolv, and the quantum task homo from QM9. The color in the visualization denotes the ground-truth discrete or continuous labels in downstream tasks. The clustering results are presented in Figure~\ref{fig:abi2}. In addition, we quantitatively calculate the Davies–Bouldin Index~\citep{davies1979cluster}, a metric employed to assess clustering results, with lower values indicating better performance, and present these values below each corresponding subfigure.

From Figure~\ref{fig:abi2}, two key insights emerge. Firstly, the first row demonstrates that our methods yield meaningful clustering results across all types of tasks, where samples with the same or similar labels are clustered together. This indicates that our pre-training tasks are neither conflicting nor redundant but rather complementary. Secondly, examining the second and third rows, it is obvious that 3D denoising is more advantageous for quantum tasks compared to biological tasks, while the masking method exhibits the opposite trend: it is helpful for biological tasks but not as effective for quantum tasks. This observation serves as a solid validation of our earlier theoretical analyses concerning various pre-training tasks.

\section{Missing Proofs and Analysis}\label{app: sec proof&analyse}
\subsection{Proof of Theorem~\ref{thm: mutual bound}}
As introduced in section~\ref{sec:theory unify methods}, we formalize the reconstruction loss and contrastive loss as follows:
\begin{definition}[Reconstruction loss]
    \begin{equation}
        \gL_{\text{RC}}=\E_{p(x)}\mathbb{E}_{p(\tilde{x}|x)}||g_\phi(h_\psi(f_\theta(\tilde{x}))) - x||_2
    \end{equation}
\end{definition}
\begin{definition}[Contrastive loss]
    \begin{equation}
        \gL_{\text{CL}}=\E_{p(x)}\mathbb{E}_{p(\tilde{x}|x)}||h_\psi(f_\theta(\tilde{x})) - \mathcal{SG}(f_\theta(x))||_2
    \end{equation}
\end{definition}
We also introduce two auxiliary losses to bridge the reconstructive and contrastive targets.
\begin{definition}[Regularization loss]
    \begin{equation}
         \gL_{\text{reg}}=\E_{p(x)} ||g_\phi( \mathcal{SG}(f_\theta(x)) ) - x||_2
    \end{equation}
\end{definition}
\begin{definition}[Modified reconstruction loss]
    \begin{equation}
        \gL_{\text{RC2}}=\E_{p(x)}\mathbb{E}_{p(\tilde{x}|x)}||g_\phi( \mathcal{SG}(f_\theta(x)) )-g_\phi(h_\psi(f_\theta(\tilde{x})))||_2
    \end{equation}
\end{definition}
The proof of Theorem~\ref{thm: mutual bound} relies on the following two lemmas.
\begin{lemma}\label{lemma1}
    The contrastive loss and the modified reconstruction loss are upper and lower bounded by each other when $\lambda_{\text{max}}$ and $\lambda_{\text{min}}$ are non-zero.
    \begin{equation}\label{eq:CL-RC2}
        \lambda_{\text{max}} \gL_{\text{CL}}\geq \gL_{\text{RC2}} \geq \lambda_{\text{min}} \gL_{\text{CL}},
    \end{equation}
     where $\lambda_{\text{max}}$ and $\lambda_{\text{min}}$ are non-negative constants defined in the proof.
\end{lemma}
\begin{proof}
    Denote vectors $a\triangleq f_\theta(x)\in \R^\dimz$ and $b\triangleq h_\psi(f_\theta(\tilde{x}))\in \R^\dimz$. Since $g_\phi$ is continuously differentiable, by the mean value formula of multivariant real-valued function~\cite{Zorich2015I}, $\forall j=1,\cdots,\dimx$, $\exists \xi_j$ in the interval of $a$ and $b$, s.t.
             $g_{\phi,j} (a)-g_{\phi,j}(b)= g'_{\phi,j}(\xi_j)^\top(a-b)$, where $g_{\phi,j}$ is the j th output dimenstion of $g_{\phi}$ and $g'_{\phi,j}(\xi_j)$ denotes the gradient with respect to $\xi_j$.
            Denote $G(x,\tilde{x})=(g'_{\phi,1}(\xi_1),\cdots,g'_{\phi,\dimx}(\xi_\dimx))^\top\in\R^{\dimx\times\dimz}$, $g_{\phi}(a)-g_{\phi}(b)= G(x,\tilde{x})(a-b)$.
    Then the modified reconstruction loss for the specific sample and its augmentation is
    \begin{equation}
        \begin{aligned}
            &l_{\text{RC2}}(x,\tilde{x})\triangleq||g_\phi( \mathcal{SG}(f_\theta(x)) )-g_\phi(h_\psi(f_\theta(\tilde{x})))||_2\\
            =&\left\{ [\mathcal{SG}(f_\theta(x))-h_\psi(f_\theta(\tilde{x}))]^\top G(x,\tilde{x})^\top G(x,\tilde{x})[\mathcal{SG}(f_\theta(x))-h_\psi(f_\theta(\tilde{x}))]\right\}^{1/2}
        \end{aligned}
    \end{equation}  
    The matrix $G^\top(x,\tilde{x}) G(x,\tilde{x})$ is symmetric semi-positive
definite matrix, thus its largest and smallest eigenvalue is non-negative and can be denoted as $\hat{\lambda}_{\text{max}}^2(x,\tilde{x})$ and $\hat{\lambda}_{\text{min}}^2(x,\tilde{x})$, where $\hat{\lambda}_{\text{max}}$ and $\hat{\lambda}_{\text{min}}$ are also non-negative. We use $l_{\text{CL}}(x,\tilde{x})$ to denote $||\mathcal{SG}(f_\theta(x))-h_\psi(f_\theta(\tilde{x}))||_2$. Then
        \begin{equation}
        \begin{aligned}
            & \lambda_{\text{min}}l_{\text{CL}}(x,\tilde{x}) \leq \hat{\lambda}_{\text{min}}(x,\tilde{x})l_{\text{CL}}(x,\tilde{x}) \leq l_{\text{RC2}}(x,\tilde{x})\leq \hat{\lambda}_{\text{max}}(x,\tilde{x})l_{\text{CL}}(x,\tilde{x}) \leq \lambda_{\text{max}}l_{\text{CL}}(x,\tilde{x}),
        \end{aligned}
    \end{equation} 
    where $\lambda_{\text{min}}=\inf_{x,\tilde{x}}\lambda_{\text{min}}(x,\tilde{x})$, $\lambda_{\text{max}}=\sup_{x,\tilde{x}}\lambda_{\text{max}}(x,\tilde{x})$ are constants independent on $x$ and $\tilde{x}$.
    After taking expectation with respect to $p(x,\tilde{x})$ to the terms in the inequality, we obtain the inequality of loss functions in \eqref{eq:CL-RC2}.
\end{proof}
\begin{lemma}\label{lemma2}
    \begin{equation}
        \gL_{\text{RC2}} + \gL_{\text{reg}} \geq \gL_{\text{RC}} \geq \gL_{\text{RC2}} - \gL_{\text{reg}}
    \end{equation}
\end{lemma}
\vspace{-10pt}
\begin{proof}
    This is a straightforward application of the triangle inequality with the $L^2$ distance.
    Denote vectors $c\triangleq g_\phi (f_\theta(x))\in\R^\dimx$, $d\triangleq g_\phi( h_\psi( f_\theta(\tilde{x})))\in\R^\dimx$. Utilizing the Minkowski inequality, we derive  $||c-d||_2+||x-c||_2\geq||(c-d)+(x-c)||_2=||x-d||_2$. This implies $\gL_{\text{RC2}} + \gL_{\text{reg}} \geq \gL_{\text{RC}}$. Analogously, we establish $\gL_{\text{RC}} + \gL_{\text{reg}} \geq \gL_{\text{RC2}}$.
\end{proof} 
\vspace{-10pt}
By employing Lemma~\ref{lemma1} and Lemma~\ref{lemma2}, we establish that  $\lambda_{\text{max}} \gL_{\text{CL}}  + \gL_{\text{reg}}\geq \gL_{\text{RC2}} + \gL_{\text{reg}} \geq \gL_{\text{RC}}$. When $\lambda_{\text{max}}>0$, dividing both sides of the inequality by $\lambda_{\text{max}}$ concludes the proof of \eqref{eq:thm CL>RC}. Conversely, we have $\gL_{\text{RC}} + \gL_{\text{reg}} \geq \gL_{\text{RC2}}\geq \lambda_{\text{min}} \gL_{\text{CL}}$, thus completing the proof of \eqref{eq:thm RC>CL}.
\vspace{-5pt}
\subsection{Discussing the Non-zero Conditions of Theorem~\ref{thm: mutual bound}}\label{app: subsec analyse}
Concerning $\lambda_{\text{max}}$, it equals zero if and only if $G(x,\tilde{x})=0$ for all $x$ and $\tilde{x}$. In this case, $g_{\phi,j} (f_\theta(x))=g_{\phi,j}(h_\psi(f_\theta(\tilde{x})))$ for all $j$, $x$ and $\tilde{x}$, indicating a collapsed decoder that maps all features to a constant vector. This case is avoided by optimizing $\gL_{\text{reg}}$ or $\gL_{\text{RC}}$.



 As for $\lambda_{\text{min}}$, it is non-zero if and only if $\dimz\leq\dimx$ and $G(x,\tilde{x})$ is full column rank for all $x$ and $\tilde{x}$. The dimension condition is usually satisfied in practice since the input of the encoder is usually high-dimensional embeddings. We can gain more insights in the linear decoder case, where $G$ is exactly the linear decoder matrix.
 Denote the feature and the reconstructed output as $z$ and $\hat{x}$ respectively. Then we have $\hat{x}=Gz=\sum_{k=1}^{\dimz}G_{:, k}z_k$, where $G_{:, k}$, $k=1,\cdots \dimz$ are the column vectors of $G$. Full column rank means $G_{:, k}$ is linearly independent, indicating there are no redundant dimensions in the feature for reconstruction. Otherwise, there exists an elementary matrix $P$ s.t. $GP$ has at least one column vector $(GP)_{:, l}$ is a zero vector. Then $(P^{-1}z)_l$ is redundant in the sense that it does not affect the reconstruction.
\subsection{Proof of Theorem~\ref{thm:culster}}\label{app sec: proof cluster}
\vspace{-5pt}
\begin{proof}
    \begin{equation}\label{eq.18}
        \begin{aligned}        &\gL_{\text{CL}}^{(\text{SimSiam})}=\E_{p(x)}\E_{p(\tilde{x}|x)}||h_\psi(f_\theta(\tilde{x})) - \mathcal{SG}(f_\theta(x))||_2^2 \mbox{  , where the representations are normalized.}\\
            &= 2-2\E_{p(x)}\E_{p(\tilde{x}|x)}f_\theta(x)^\top h_\psi(f_\theta(\tilde{x}))\\ 
             &=2-\E_{p(x)}2f_\theta(x)^\top \left( \int p(\tilde{x}|x)h_\psi(f_\theta(\tilde{x})) \dd \tilde{x}\right)\\
             &\geq 2-\E_{p(x)}\left[||f_\theta(x)||_2^2+\left( \int p(\tilde{x}|x)h_\psi(f_\theta(\tilde{x})) \dd \tilde{x}\right)^\top \left( \int p(\tilde{x}|x)h_\psi(f_\theta(\tilde{x})) \dd \tilde{x}\right)\right]\\
             &=1-\int \int\E_{p(x)}  p(\tilde{x}|x)p(\tilde{x}'|x)h_\psi(f_\theta(\tilde{x})) ^\top h_\psi(f_\theta(\tilde{x}')) \dd \tilde{x} \dd \tilde{x}'.
        \end{aligned}
    \end{equation}
    Denote $\omega(\tilde{x},\tilde{x}')\triangleq\E_{p(x)}p(\tilde{x}|x)p(\tilde{x}'|x)$ to establish a probability of generating two augmentations $\tilde{x}$ and $\tilde{x}'$ from the same input, and it is normalized $\int \int \omega(\tilde{x},\tilde{x}')\dd \tilde{x} \dd \tilde{x}'= \E_{p(x)} \int  p(\tilde{x}|x) \dd \tilde{x} \int p(\tilde{x}'|x)\dd \tilde{x}'=1 $. A similar concept is initially introduced in \cite{haochen2021spectralcl} as the weight of the augmentation graph. 
    \begin{equation}
    \begin{aligned}
        \gL_{\text{CL}}^{(\text{SimSiam})}&\geq 1-\E_{\omega(\tilde{x},\tilde{x}')} h_\psi(f_\theta(\tilde{x})) ^\top h_\psi(f_\theta(\tilde{x}'))\\
         &=\frac{1}{2} \E_{\omega(\tilde{x},\tilde{x}')}  ||h_\psi(f_\theta(\tilde{x}))- h_\psi(f_\theta(\tilde{x}'))||_2^2\triangleq \frac{1}{2}\gL_{symm}
    \end{aligned}         
    \end{equation}
    Therefore, the SimSiam loss can serve as an upper bound for the symmetric contrastive loss that aligns the positive pairs. Then we convert the pairwise alignment into the centered clustering formulation.
    \begin{equation}
    \begin{aligned}
    \gL_{\text{cluster}} &= \E_{p(x)}\E_{p(\tilde{x}|x)} ||h_\psi(f_\theta(\tilde{x}))- \E_{p(\tilde{x}|x)} h_\psi(f_\theta(\tilde{x}))||_2^2\\
        =&\E_{p(x)} \left( \E_{p(\tilde{x}|x)} ||h_\psi(f_\theta(\tilde{x}))||_2^2- \left(\E_{p(\tilde{x}|x)} h_\psi(f_\theta(\tilde{x}))\right)^\top \left(\E_{p(\tilde{x}|x)} h_\psi(f_\theta(\tilde{x}))\right)\right)\\
        =&\frac{1}{2}\E_{p(x)}\E_{p(\tilde{x}|x)}\E_{p(\tilde{x}'|x)}  ||h_\psi(f_\theta(\tilde{x}))- h_\psi(f_\theta(\tilde{x}'))||_2^2\\
       =&\frac{1}{2}\E_{\omega(\tilde{x},\tilde{x}')}  ||h_\psi(f_\theta(\tilde{x}))- h_\psi(f_\theta(\tilde{x}'))||_2^2
        =\frac{1}{2}\gL_{symm}\leq \gL_{\text{CL}}^{(\text{SimSiam})}
    \end{aligned}         
    \end{equation}
\end{proof}
\subsection{Corollary of Theorem~\ref{thm:culster}}\label{app sec corollary}
\begin{corollary}[ Relations between cross-modal contrastive learning and clustering]
Theorem~\ref{thm:culster} still holds when the input and its augmentations employ distinct encoders or parameters, for example in the case of cross-modal contrastive learning. 
    \begin{equation}
    \begin{aligned}
       \gL_{\text{CL,cross-modal}}^{(\text{SimSiam})}\triangleq\E_{p(x)}\E_{p(\tilde{x}|x)}||h_\psi(f_\theta(\tilde{x})) - \mathcal{SG}(f'_{\theta'}(x))||_2^2\geq \gL_{\text{cluster}}, 
    \end{aligned}
    \end{equation}
     where the representations are defined to be normalized: $||f'_{\theta'}(\cdot)||_2^2=||h_\psi(f_\theta(\cdot))||_2^2=1$. $\gL_{\text{cluster}}  \triangleq \E_{p(x)}\E_{p(\tilde{x}|x)} $ $||h_\psi(f_\theta(\tilde{x}))- \E_{p(\tilde{x}|x)} h_\psi(f_\theta(\tilde{x}))||^2_2$ describes the mean distance between the samples in the cluster and the cluster center in the representation space of augmentations.
\end{corollary}
\begin{proof}
The heterogeneous encoders do not alter the result of \eqref{eq.18} since the encoder of the input $f'_{\theta'}$ will diminish as long as the representation is normalized.
\begin{equation}
        \begin{aligned} 
         &\gL_{\text{CL,cross-modal}}^{(\text{SimSiam})}=\E_{p(x)}\E_{p(\tilde{x}|x)}||h_\psi(f_\theta(\tilde{x})) - \mathcal{SG}(f'_{\theta'}(x))||_2^2 \\
            &= 2-2\E_{p(x)}\E_{p(\tilde{x}|x)}f'_{\theta'}(x)^\top h_\psi(f_\theta(\tilde{x}))\\ 
             &=2-\E_{p(x)}2f'_{\theta'}(x)^\top \left( \int p(\tilde{x}|x)h_\psi(f_\theta(\tilde{x})) \dd \tilde{x}\right)\\
             &\geq 2-\E_{p(x)}\left[||f'_{\theta'}(x)||_2^2+\left( \int p(\tilde{x}|x)h_\psi(f_\theta(\tilde{x})) \dd \tilde{x}\right)^\top \left( \int p(\tilde{x}|x)h_\psi(f_\theta(\tilde{x})) \dd \tilde{x}\right)\right]\\
             &=1-\int \int\E_{p(x)}  p(\tilde{x}|x)p(\tilde{x}'|x)h_\psi(f_\theta(\tilde{x})) ^\top h_\psi(f_\theta(\tilde{x}')) \dd \tilde{x} \dd \tilde{x}'.
        \end{aligned}
    \end{equation}
Hence, we reach an identical result as expressed in Equation \eqref{eq.18}. The remaining proof aligns precisely with those presented in the proof of Theorem~\ref{thm:culster}.
\end{proof}

\section{Experimental Details}
\subsection{Dataset Description}\label{sec:app data}
QM9~\citep{ramakrishnan2014quantum, ruddigkeit2012enumeration} is a quantum chemistry dataset that offers a single equilibrium conformation along with 12 labels covering geometric, energetic, electronic, and thermodynamic properties for 134,000 stable small organic molecules comprised of CHONF atoms. The dataset is split following typical settings, resulting in a training set of 110,000 samples, a validation set of 10,000 samples, and a test set containing the remaining 10,831 samples.

MD17~\citep{chmiela2017machine} comprises molecular dynamics trajectories for 8 small organic molecules. Each molecule in the dataset is associated with 150k to almost 1M conformations, and includes total energy and force labels. Our focus lies on the challenging task of force prediction. In accordance with a standard limited data setting, the model undergoes training on a subset of 1000 samples, with 50 allocated for validation, while the remaining data is employed for testing.

MD22~\citep{Chmiela2023MD22} consists of molecular dynamics trajectories covering 7 molecules from four major classes of biomolecules and supramolecules, ranging from a 42-atom peptide to a double-walled nanotube containing 370 atoms. The dataset provides labels for both total energy and force. Our focus lies on the challenging task of force prediction. The dataset is split by a ratio of 8:1:1 into the train, validation, and test sets.

MoleculeNet~\citep{wu2018moleculenet} is widely recognized as a benchmark for predicting a range of molecular properties. 
It covers biological tasks such as BBBP, SIDER, ClinTox, Tox21, Toxcast, BACE, HIV, and MUV, as well as physicochemical challenges, including FreeSolv, ESOL, and Lipophilicity.
Each dataset is split into training, validation, and test sets following an 8:1:1 ratio, implemented through the scaffold method. 

\subsection{Construction of Hierarchical Data}\label{sec: hierarchical construction}
The construction of the dataset utilized in~\ref{sec:hiervis} begins with the selection of five common heterocyclic scaffolds, each possessing different chemical properties but exhibiting a similar structure to make the clustering more difficult. Subsequently, we randomly sample 128 molecules for each scaffold and conduct molecular simulations on each molecule using OpenMM~\citep{eastman2013openmm}, generating four molecular dynamics trajectories. From each trajectory, we randomly select four adjacent conformations, resulting in approximately 16 conformations per molecule. In total, this meticulously constructed dataset consists of around 10,000 conformations.

\subsection{Hyperparameter Settings}\label{app: setting}
\begin{table}[h!]
    \caption{Hyperparameters for pre-training. }
    \label{table:app setting pretrain}
    \vskip 0.15in
    \begin{center}
    \begin{footnotesize}
    \begin{tabular}{lc}
    \toprule
    Parameter & Value or description\\
     \midrule 
   Batch size &  256\\
Optimizer  & 	AdamW	\\
Adam betas & (0.9, 0.999)\\
Max Learning rate & 	0.0004	\\
Warm up steps & 	10000	\\
Learning rate decay policy	 & Cosine\\
	Learning rate factor & 	0.8\\
 Training steps & 1500000 \\
 \midrule 
 3D encoder layers number & 8 \\
 3D encoder attention head number & 8 \\
 3D encoder embedding dimension & 256 \\
 \midrule 
 2D encoder layers number & 12\\
 2D encoder attention head number & 32 \\
 2D encoder embedding dimension & 512 \\
 
   \bottomrule
    \end{tabular}
    \end{footnotesize}
    \end{center}
    \vskip -0.1in
\end{table}

In our pre-training stage, we employ TorchMD-NET~\citep{tholke2022torchmd} as the 3D encoder backbone and Graphormer~\citep{ying2021transformers} as the 2D encoder. 
  The loss weights for fragment masking loss, denoising loss, and cross-modal distillation loss are set at a ratio of 1:1:1. Within the masking task, the masking ratio is configured to be 0.2. For the denoising task, the standard deviations of torsion Gaussian noise and coordinate Gaussian noise are set to 2 and 0.04, respectively. The temperature $\tau$ is set to 0.5 for the cross-modal distillation task. Additional hyperparameters associated with the network structure and pre-training process can be found in Table~\ref{table:app setting pretrain}.

In line with previous methods, we employ grid search to find the optimal hyperparameters for tasks in MoleculeNet. The specific search space for each task is detailed in Table~\ref{table:app setting molnet}.

\begin{table*}[t]
\setlength{\tabcolsep}{4pt}
\caption{Search space for MoleculeNet dataset, where [...] represents continuous interval, $\{...\}$ denotes discrete candidate values.}
\label{table:app setting molnet}
\vskip 0.15in
\begin{center}
\begin{small}
\begin{tabular}{lllll}
\hline
Parameter&Classification tasks&Regression tasks\\\hline
Learning rate&{[}1e-4,1e-3{]}&{[}1e-4,1e-3{]}\\
Batch size&\{16,32,64\}&\{8,16,32\}\\
Epochs&\{5,10,25,50\}&\{25,50\}\\
Weight decay&\{0,1e-5\}&\{0,1e-5\}\\\hline
\end{tabular}
\end{small}
\end{center}
\vskip -0.1in
\end{table*}

\begin{table}[t]
    \caption{Hyperparameters for fine-tuning on QM9. }
    \label{table:app setting qm9}
    \vskip 0.15in
    \begin{center}
    \begin{footnotesize}
    \begin{tabular}{lc}
    \toprule
    Parameter & Value or description\\
     \midrule  
Train/Val/Test Splitting	 & 110000/10000/remaining data	\\
Batch size & 	128	\\
  \midrule  
Optimizer	 & AdamW	\\
Warm up steps & 	10000	\\
Max Learning rate	 & 0.0004	\\
Learning rate decay policy & 	Cosine	\\
	Learning rate factor	 & 0.8\\
	Cosine cycle length	 & 300000 (500000 for tasks $\alpha$, $ZPVE$, $U_0$, $U$, $H$, $G$) \\
   \bottomrule
    \end{tabular}
    \end{footnotesize}
    \end{center}
    \vskip -0.1in
\end{table}

\begin{table}[t]
    \caption{Hyperparameters for fine-tuning on MD17. }
    \label{table:app setting md17}
    \vskip 0.15in
    \begin{center}
    \begin{footnotesize}
    \begin{tabular}{lc}
    \toprule
    Parameter & Value or description\\
     \midrule  
  Train/Val/Test Splitting &	950/50/remaining data	\\
  Batch size &	8	\\
  \midrule  
Optimizer&	AdamW	\\
Warm up steps	&1000	\\
Max Learning rate	&0.0005	\\
Learning rate decay policy&	ReduceLROnPlateau (Reduce Learning Rate on Plateau) scheduler	\\
	Learning rate factor&	0.8\\
	Patience	&30\\
	Min learning rate	&1.00E-07\\
\midrule
Force weight	&0.8		\\
Energy weight	&0.2		\\
   \bottomrule
    \end{tabular}
    \end{footnotesize}
    \end{center}
    \vskip -0.1in
\end{table}

\begin{table}[h!]
    \caption{Hyperparameters for fine-tuning on MD22. }
    \label{table:app setting md22}
    \vskip 0.15in
    \begin{center}
    \begin{footnotesize}
    \begin{tabular}{lc}
    \toprule
    Parameter & Value or description\\
     \midrule  
Train/Val/Test Splitting & 80\%/10\%/10\% split same with \cite{coordnoneq2023} 	\\
  Batch size &	\makecell[c]{4 for double-walled-nanotube\\16 for Ac-Ala3-NHMe and DHA \\ 8 for the other 4 tasks}	\\
  \midrule  
Optimizer&	AdamW	\\
Epochs &50 \\
Max Learning rate	& 0.001\\
Learning rate decay policy& Cosine\\
Warm up steps	& 30\% steps of the first training epoch	\\
Patience steps	& 70\% steps of the first training epoch	\\
	Min learning rate	&1.00E-07\\		
 \midrule  
Force weight	&0.8		\\
Energy weight	&0.2		\\
   \bottomrule
    \end{tabular}
    \end{footnotesize}
    \end{center}
    \vskip -0.1in
\end{table}
The hyperparameters for fine-tuning on QM9, MD17, and MD22 are outlined in Tables \ref{table:app setting qm9}, \ref{table:app setting md17} and \ref{table:app setting md22}. When performing fine-tuning, we incorporate the Noisy Node task, following the approach in \citep{godwin2021simple,pmlr-v202-feng23c}. The loss weight for this task is set to 0.1.

\subsection{The Impact of Data Accuracy and Diversity}
Both the accuracy and diversity of 3D pre-training data have an impact on downstream task results. While we do not have a definitive conclusion, we have observed some phenomena that offer valuable insights for the community.

From the perspective of accuracy and diversity of 3D conformers, our pretraining data can be categorized into two distinct types. The first type comprises data sourced from the PubChemQC Project~\cite{nakata2017pubchemqc}, where conformations are calculated using the DFT method, ensuring high quality. However, due to the substantial computational costs involved, this dataset contains a limited number and variety of molecules, resulting in less diversity. The second type consists of data collected from various sources via the Uni-Mol~\cite{zhou2023unimol}, utilizing RDKit to generate conformers more efficiently. Consequently, this dataset encompasses a broader range of molecules, enhancing diversity.

Through comprehensive experiment and analysis, we conclude that the denoising task prefers accurate conformation. Conversely, for masking and contrastive learning tasks, the diversity of the dataset plays a more critical role. As demonstrated by previous 3D denoising methods~\citep{zaidi2022pre, feng2023fractional}, the denoising task relies on highly accurate equilibrium conformations as input. As observed in table~\ref{table:accuracy},  pre-training UniCorn solely with DFT-calculated data achieves comparable performance on QM9 compared with UniCorn which uses additional RDKit data. Since quantum tasks such as QM9 and MD17 primarily involve small and simple molecules, the accuracy of the data may be more important than its diversity in the denoising task.
\begin{table}[h!]
    \caption{Performance (MAE, $\downarrow$) on QM9 quantum tasks. The best results are in bold.}
    \label{table:accuracy}
    \vskip 0.15in
    \begin{center}
    \begin{footnotesize}
    \begin{tabular}{lcccc}
    \toprule
 {QM9 	}&{\makecell[c]{$\epsilon_{HOMO}$ (meV)}   }&{\makecell[c]{$\epsilon_{LOMO}$ (meV)}}&{ \makecell[c]{$\Delta\epsilon$(meV)}  }&{	AVG}\\
     \midrule  
{UniCorn (DFT)	}&{13.2	}&{12	}&{25.1	}&{16.7}\\
{UniCorn (DFT + RDkit)	}&{\textbf{13}	}&{\textbf{11.9}	}&{\textbf{24.9}	}&{\textbf{16.6}}\\
   \bottomrule
    \end{tabular}
    \end{footnotesize}
    \end{center}
    \vskip -0.1in
\end{table}
Data diversity is more important in the context of masking and 2D-3D contrastive learning pre-training tasks, because their corresponding downstream tasks in MoleculeNet contain more complex and diverse molecules covering a border range of biological properties that don't need highly accurate conformation. Moreover, more diverse 3D conformations of one molecule are also beneficial for 2D-3D contrastive learning as demonstrated by GraphMVP and 3D InfoMax. As depicted in Table~\ref{table:diversity},  reveals that UniCorn pre-trained with additional RDKit data outperforms UniCorn solely pre-trained with DFT data on biological and physicochemical tasks.
\begin{table}[h!]
    \caption{Performance (ROC-AUC\%, $\uparrow$; RMSE, $\downarrow$) on MoleculeNet tasks. The best results are in bold.}
    \label{table:diversity}
    \vskip 0.15in
    \begin{center}
    \begin{footnotesize}
    \begin{tabular}{lcccccccc}
    \toprule
{Models }&{	BBBP↑}&{	BACE↑	}&{ClinTox↑	}&{Tox21↑	}&{ToxCast↑	}&{SIDER↑}&{	FreeSolv↓}&{	Lipo↓}\\
   \midrule  
{UniCorn(DFT)	}&{73.0(0.9)}&{	83.6(0.2)	}&{88.5(1.4)	}&{79.2(1.8)}&{	68.6(0.3)	}&{63.1(0.7)}&{	1.831(0.189)}&{	0.589(0.008)}\\
{UniCorn(DFT + RDKit)}&{	\textbf{74.2(1.1)}}&{	\textbf{85.8(1.2)}	}&{\textbf{92.1(0.4)}}&{	\textbf{79.3(0.5)}}&{	\textbf{69.4(1.1)}	}&{\textbf{64.0(1.8)}	}&{\textbf{1.555(0.075)}	}&{\textbf{0.591(0.016)}}\\
   \bottomrule
    \end{tabular}
    \end{footnotesize}
    \end{center}
    \vskip -0.1in
\end{table}
\section{Related Work}\label{app: Related work}
\subsection{Unified Molecular Pre-training Methods}
Recently, several unified pre-training methods have been proposed to be adopted for a diverse range of molecular tasks. \citet{zhou2023unimol} and \citet{fang2022geometry} develop pre-training tasks centered around the molecular 3D structure. involving the prediction of geometry information and 3D denoising. Other multimodal pre-training methods focus on integrating molecular 2D graphs and 3D structures. \citet{luo2022one} designs a versatile Transformer structure capable of accommodating both 2D and 3D inputs, engages in denoising and quantum property prediction as pre-training tasks. \citet{liu2023group} and \citet{li2023multimodal} take a unique approach by incorporating molecular 2D and 3D information into the diffusion process for representation learning. \citet{yu2023unified} merges atom relations involving different modalities, into a unified relation matrix, then predicts 2D graph and 3D structure information separately. However, previous methods have fallen short in comprehending the in-depth relationship between pre-training methods and downstream tasks, and struggle to achieve satisfactory results across both microscopic and macroscopic tasks.

\subsection{Unifying Masking and Contrastive Learning in Other Fields}
 Recently, some theoretical works have studied masking and contrastive learning in other fields. In the field of computer vision, \citet{zhang2022theoryMAE} investigates in the mask image modeling setting and lower bound the masking loss by a type of contrastive loss that aligns the mask-induced positive pairs.  \citet{qi2023recon} provides a unified view of masking and contrastive learning through the student-teacher paradigm. 
 In the field of graph representation learning, \citet{maskgraph} analyzes masking graph modeling in an information-theoretic view. They first prove that graph autoencoder and contrastive learning have asymptotic equivalent solutions and then explain the benefit of graph masking by reducing redundancy between views. 

\section{Pseudocode for Pre-training and Fine-tuning Algorithms}\label{sec:pseodocode}

We provide Algorithm \ref{alg:unicorn} and Algorithm \ref{alg:finetune} presenting the pseudocode for the pre-training and fine-tuning processes of UniCorn, respectively.

\begin{algorithm}[h!]
\caption{Applying UniCorn to pre-training}\label{alg:unicorn}
\begin{algorithmic}[1]
\Require
\Statex$f^{2d}$: 2D encoder
\Statex$f^{3d}$: 3D encoder
\Statex$g_m$: MLP head for 2D masking
\Statex $g_d$: MLP head for 3D denoising
\Statex $h_p$: MLP head for feature alignment between 3D and 2D representation
\Statex$\mathcal{G}$: 2D molecular graph
\Statex$\mathcal{C}=(\mathcal{V}, \mathcal{X})$: 3D molecular conformation

\Statex$m$: Mask ratio of 2D fragment masking

\Statex$X$: Unlabeled pre-training dataset
\Statex$x_i$: Input sample
\Statex$T$: Training steps
\While{$T \neq 0$}
    \State $x_i$ = dataloader($X$), $x_i$=$(\gG, \gC)$  \Comment{random sample $x_i$ from $X$, $x_i$ contains 2D molecular graph and 3D conformation}
    \State According to the mask ratio $m$, randomly mask a specific number of fragments denoted as $\mathbf{s}_{m}$ in $\gG$, resulting in the masked 2d graph $\mathcal{G}_{\backslash m}$
    
    \State $\mathcal{L}_m = -\mathbb{E}_{\mathcal{G},\mathcal{G}_{\backslash m}}\sum_{\mathbf{s}_{m} } \log \left[p_{(g_m, f^{2d})}\left(\mathbf{s}_{m} |\mathcal{G}_{\backslash m} \right)\right]$ \Comment{Calculate the masked fragment loss}
    
    \State Introduce torsion Gaussian noise to $\gC$ to obtain $\gC_{a} = (\mathcal{V}, \mathcal{X}_a)$, then add the coordinate Gaussian noise to $\gC_{a}$ to derive $\tilde{\gC} = (\mathcal{V}, \tilde{\mathcal{X}})$

    \State $\mathcal{L}_{dn} = \mathbb{E}_{\tilde{\mathcal{C}},\mathcal{C}_a}||g_{d}(f^{3d}(\tilde{\mathcal{C}})) - (\tilde{\mathcal{X}} - \mathcal{X}_a)||_{2}^{2}$ \Comment{Calculate the 3D denoising loss}
    
    \State $\mathcal{L}_{cl} = - \mathbb{E}_{\mathcal{G}, \tilde{\mathcal{C}}}[ \log \frac{e^{\text{cos}(\boldsymbol{z}^{g}, \boldsymbol{z}^{c}) / \tau}}{ \sum\limits_{\boldsymbol{z}_j^{c}} e^{\text{cos}(\boldsymbol{z}^{g}, \boldsymbol{z}_j^{c}) / \tau}} +   \log \frac{e^{\text{cos}(\boldsymbol{z}^{c}, \boldsymbol{z}^{g}) / \tau}}{ \sum\limits_{\boldsymbol{z}_j^{g}} e^{\text{cos}(\boldsymbol{z}^{c}, \boldsymbol{z}_j^{g}) / \tau}}]$, where $\boldsymbol{z}^{c} = h_p ( f^{3d} ( \tilde{\mathcal{C}} ))$, $\boldsymbol{z}^{g} = f^{2d} (\mathcal{G})$, $\boldsymbol{z}_j^{c}\in \mathbb{N}^c\cup\boldsymbol{z}^{c}$ and $ \boldsymbol{z}_j^{g} \in \mathbb{N}^g\cup\boldsymbol{z}^{g}$, $\mathbb{N}^c$ and $\mathbb{N}^g$ denotes in-batch negative samples of 2D graphs and 3D conformations \Comment{Calculate the Cross-modal contrastive loss}


    \State Loss = $\mathcal{L}_{dn} + \mathcal{L}_{m} + \mathcal{L}_{cl}$
    \State Optimise(Loss)
    \State $T = T - 1$
\EndWhile
\end{algorithmic}
\end{algorithm}

\begin{algorithm}[h!]
\caption{Applying UniCorn to fine-tuning}\label{alg:finetune}
\begin{algorithmic}[1]
\Require
\Statex $f^{3d}$:Pre-trained 3D encoder
\Statex $h_p$: MLP head for property prediction
\Statex $h_d$: MLP head for Noisy Nodes
\Statex $\mathcal{C}=(\mathcal{V}, \mathcal{X})$: 3D molecular conformation 
\Statex$X$: Training dataset
\Statex$x_i$: Input sample
\Statex$y_i$: Label of $x_i$
\Statex$\Delta{x_i} \sim \mathcal{N}(0, {\tau}^2I_{3N})$, $N$ is atom number of $x_i$
\Statex$T$: Training steps
\Statex$task$: the current fine-tuning downstream task, which may belong to MoleculeNet, QM9, MD17, or MD22
\Statex$\lambda_{n}$: Loss weight of Noisy Nodes
\While{$T \neq 0$}
    \State $x_i, y_i$ = dataloader($X$), $x_i$=$(\mathcal{C})$  \Comment{random sample $x_i$ and corresponding label $y_i$ from $X$}

    \State ${y_{i}^{pred}= h_p(f^{3d}(\tilde{x}_i))}^{\ast}$
    \State Loss = PropertyPredictionLoss$(y_{i}^{pred}, y_i)$
    \If{$task$ in {QM9, MD17, MD22}}
        \State ${L_d=\lambda_{n}||h_d(f^{3d}(\tilde{x}_i)) - \Delta{x}_i||_{2}^{2}}^{\ast}$ \Comment{Calculate loss of Noisy Nodes} 
        \State Loss += $\mathcal{L}_d$
    \EndIf
    \State Optimise(Loss)
    \State $T = T - 1$
\EndWhile
\Statex*: For MoleculeNet, $\tilde{x}_i=x_i$. For QM9, MD17 and MD22, we apply Noisy Nodes regularization and the definition of $\tilde{x}_i$ follows~\cite {feng2023fractional}.
\end{algorithmic}
\end{algorithm}


\section{The Relationship Between Pre-training Methods and Data Modalities}
In this section, we discuss the relationship between pre-training methods and data modalities, although we did not explicitly refer to this in the main text. We select data modalities that are best suited for these utilized pre-training methods and focus primarily on the relationship between pre-training methods and downstream tasks. However, understanding the connection between pre-training methods and data modalities is crucial for comprehensively understanding the proposed methods.

Different pre-training strategies have certain preferences for specific modalities. For instance, 
previous 3D denoising methods~\cite{zaidi2022pre,feng2023fractional,ni2024sliced} are tailored for 3D conformations, as they introduce different types of noise to the 3D structure and predict the noise aiming to approximately learn the molecular force field. In contrast, masking and contrastive learning show a slight preference for certain modalities but are generally more versatile. While masking is typically conducted on the 2D graph by masking the atoms, edges, and fragments~\cite{hu2020strategies,xia2022mole,feng2023unimap}, it can also applied to masking atoms within 3D structures~\cite{zhou2023unimol,zaidi2022pre} and masking tokens for 1D SMILES sequence \cite{chithrananda2020chemberta,wang2019smiles,zhang2021mg}. Contrastive learning is particularly flexible, as it can generally align different modalities. In addition to aligning 2D graphs with 3D conformations, it can also be used to align molecules with other relevant modalities such as textual descriptions \cite{liu2023multi} or binding proteins \cite{gao2022cosp}. In our analysis, we implicitly take into account the preference for modality, focusing on one of the most common scenarios: denoising for 3D modality, masking for 2D modality, and contrastive learning for cross-modality in the main text.

\section{Novelty and Limitations}
It's essential to note that the primary novelty of this paper lies in our combination of three prevalent existing methods under a unified contrastive learning framework for pursuing a universal molecular model. Through rigorous theoretical proof and domain-expertise-driven analysis, we innovatively establish the necessity of this combination to achieve a universal representation applicable to all types of downstream tasks—an objective that previous works have failed to attain. In particular, we demonstrate for the first time that existing methods are naturally suited to different downstream tasks, shedding light on why current literature is limited to their specific preferred task. Finally, through exhaustive experiments, our approach marks the first work in the field of molecular representation learning that achieves SOTA results on tasks that span a wide range of quantum, physicochemical, and biological domains. 

We anticipate potential limitations with UniCorn concerning its data requirements. Firstly, as UniCorn aims to capture multi-grained molecular representation and involves denoising, it relies on 3D equilibrium structures as pre-training inputs, which are relatively scarcer than 2D molecular data in existing datasets. Secondly, since the 3D encoder is utilized for various downstream tasks, the necessity of 3D conformation as input may require additional 3D generation during processing. Thirdly, the accuracy of 3D inputs or failure of 3D generation could impact the outcomes of downstream tasks. However, as 3D datasets expand and 3D generation capabilities improve, we expect the limitation of data requirements for our approach as well as similar previous works like Uni-Mol~\cite{zhou2023unimol}, GEM~\cite{fang2022geometry}, and MoleBLEND~\cite{yu2023unified}, to be mitigated in the future.

\end{document}